\documentclass[pra,aps,amssymb,twocolumn,showpacs,hyperref,superscriptaddress]{revtex4}
\usepackage{color}
\usepackage{graphicx}
\usepackage{hyperref}
\usepackage{amsmath}
\usepackage{latexsym}
\usepackage{amssymb}
\usepackage{mathrsfs}
\usepackage{mathtools}
\usepackage{layout}
\usepackage{verbatim}
\usepackage{bm}
\usepackage{amsfonts,epsfig}

\begin{document}
\title{Entanglement of neutral-atom chains by spin-exchange Rydberg interaction}
\date{\today}
\author{Xiao-Feng Shi}
\affiliation{School of Physics, Georgia Institute of Technology, Atlanta, GA, 30332-0430, USA}
\author{F. Bariani}
\affiliation{Department of Physics, College of Optical Sciences and B2 Institute, University of Arizona, Tucson, Arizona 85721, USA}
\author{T. A. B. Kennedy}
\affiliation{School of Physics, Georgia Institute of Technology, Atlanta, GA, 30332-0430, USA}

\begin{abstract}
Conditions to achieve an unusually strong Rydberg spin-exchange interaction are investigated and proposed as a means to generate pairwise entanglement and realize a SWAP-like quantum gate for neutral atoms. Ground-state entanglement is created by mapping entangled Rydberg states to ground states using optical techniques. A protocol involving SWAP gate and pairwise entanglement operations is predicted to create global entanglement of a chain of $N$ atoms in a time that is independent of $N$.

\end{abstract}
%
\maketitle

\section{introduction}
Entanglement is a property of multiparticle quantum states that is essential for implementing quantum information or computation protocols~\cite{PhysRev.47.777,Nielsen2000}. As a result, schemes for the fast and efficient generation of entanglement among many quantum systems are the subject of intense theoretical and experimental efforts.
Atoms offer an ideal arena for the demonstration of quantum protocols given the stability of their ground states and the powerful optical and trapping techniques that have been developed to control their internal and external degrees of freedom \cite{Adamsreview}.
The excitation of Alkali atoms to energy levels of large principal quantum number, generically named Rydberg states, provides a way to enhance by several orders of magnitude otherwise weak neutral atom interactions.
The resulting dipole-dipole or van der Waals interaction between two highly excited Rydberg atoms allows the creation of stable entangled states of atoms in their ground electronic level~\cite{PhysRevLett.100.170504, PhysRevLett.102.240502, PhysRevA.81.052329,  Saffman2010,Isenhower2010,Urban2009,Zhang2012,Carr2013,BhaktavatsalaRao2013}. The methods employed to date \cite{PhysRevLett.100.170504, PhysRevLett.102.240502, PhysRevA.81.052329,  Saffman2010,Isenhower2010,Urban2009} rely on the Rydberg blockade effect, in which two-atom energy levels with two Rydberg excitations experience a large interaction induced shift, while energy levels with a single Rydberg excitation are unperturbed. As a consequence, under conditions in which a single Rydberg excitation is resonantly excited, the pair excitation probability is strongly suppressed~\cite{PhysRevLett.93.063001,PhysRevLett.99.163601, PhysRevLett.99.073002,PhysRevLett.107.103001, Dudin02,NPh01,PhysRevLett.109.053002, PhysRevLett.112.183002}.

The excitation of two atoms into Rydberg $S$-orbitals with different principal quantum numbers, $n_{\text{\tiny{A}}}$ and $n_{\text{\tiny{B}}}$, and opposite electron spin orientation, produces not only the Rydberg blockade shift \cite{VanDitzhuijzen2008,Han2009,Bariani2012, Gunter2013, Gorniaczyk2014,Tiarks2014,Li2014,Paredes-Barato2014}, but also a coupling that exchanges the electron spin states. While the blockade shift is usually the dominant effect, the spin-exchange coupling can be made almost equally strong with the right choice of $n_{\text{\tiny{A}}}$ and $n_{\text{\tiny{B}}}$. In this regime, when nearby atoms are optically driven, the probability to create double Rydberg excitations can be as large as one, in sharp contrast to the case when $n_{\text{\tiny{A}}}= n_{\text{\tiny{B}}}$. Furthermore, the two-atom Rydberg state created by this mechanism is one of two entangled Bell states, the triplet denoted $|r_+\rangle$, in the subspace of the two-atom Rydberg excitations. The orthogonal Rydberg Bell state, the singlet $|r_-\rangle,$ experiences a strong level shift and is effectively decoupled from the excitation. The entangled Rydberg state created in this way is metastable, but the entanglement can be mapped optically to the ground state in a time short compared to the metastable state lifetime.
In the following, we explain how to produce ground state entanglement of a pair of atoms by a sequence of three pulses, and then discuss a protocol for global entanglement of a chain of atoms involving a sequence of twelve pulses which minimizes the blockade shifts due to multiple Rydberg excitations along the chain.

\begin{figure}
\includegraphics[width=3.3in]
{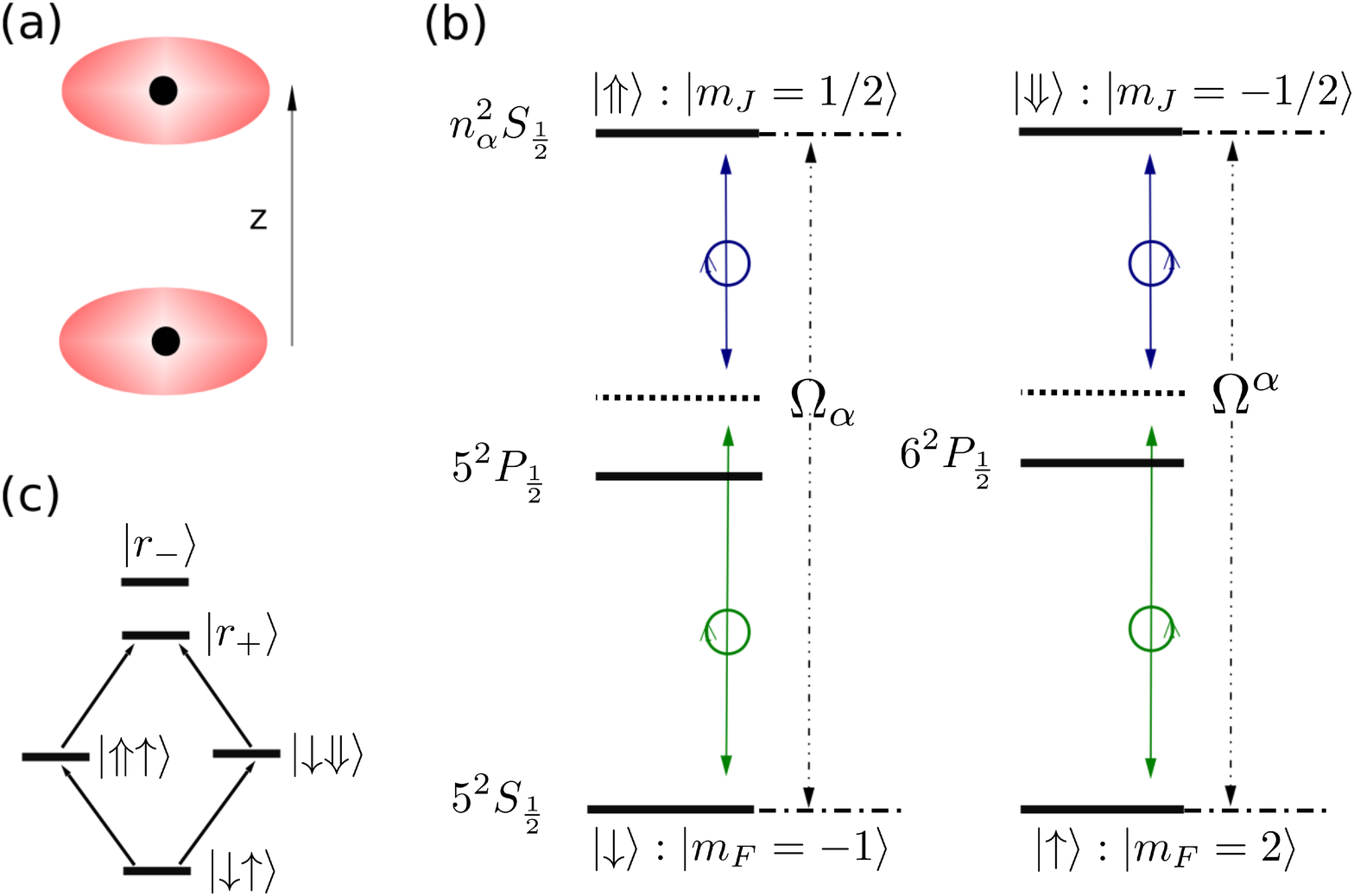}
 \caption{(Color online) Schematic illustration for generating entanglement between two $^{87}$Rb atoms. (a) Geometry considered with atoms (black dots) in tightly focused optical traps (shaded regions) along the quantization axis $\mathbf{z}$. (b) Laser excitation of atom $\alpha=A,B$ by two-photon transitions. The two-photon transition via the 5P~(6P) state of atom $\alpha$ has an effective Rabi frequency $\Omega_{\alpha}$($~\Omega^{\alpha}$). (c) Relevant levels in the excitation of Rydberg Bell state $|r_+\rangle$. \label{fig001}   }
\end{figure}

The main body of the paper gives an account of the novel interaction mechanism and its applications to quantum computation. Section II introduces the spin-exchange interaction and its potential for two-atom entanglement. Section III discusses an example to achieve pairwise entanglement via optical pumping. In Sec.~IV, we investigate the pairwise entanglement efficiency by numerical simulation. In Sec.~V, we study a quantum gate that is similar to the SWAP gate, and introduce a protocol of global entanglement of atoms in a chain. Section VI gives a summary. Additional details of the theory are given in the appendices.

\section{Two-atom entanglement}

 Consider two $^{87}$Rb atoms, denoted $A$ and $B$, respectively, which are loaded into two far-off-resonant traps created by tightly focused laser beams. The interatomic axis is defined by the vector $\mathbf{z}$. Each of the atoms can be independently driven by laser light, as shown in Fig.~\ref{fig001}.
We suppose atom $A$ is prepared in a hyperfine level of the ground state manifold identified as ``spin-up" while atom $B$ is prepared in the state ``spin-down'': $|$$\uparrow\rangle_{\text{A}}$($|$$\downarrow\rangle_{\text{B}}) \equiv |5^2S_{\frac{1}{2}}, F = 2, m_F = 2(-1)\rangle$. These states may be optically coupled via two-photon transitions to the Rydberg states $|$$\Downarrow\rangle_{\text{A}}$($|$$\Uparrow\rangle_{\text{B}}) \equiv |n_{\text{\tiny{A(B)}}}~  ^2S_{\frac{1}{2}}, m_J = \mp 1/2 \rangle$ of atoms $A$ and $B$, respectively.
Here we use the hyperfine notation to label ground states and the fine structure notation for Rydberg levels, according to the spectroscopic resolution usually achieved in experiments~(see Appendix~\ref{appA}).

Consider two Rydberg atoms prepared in the states $|$$\Uparrow \rangle_{\text{A}}$ and $|$$\Downarrow\rangle_{\text{B}}$, respectively, and separated by a distance $L$ large enough that the states are coupled by the $C_6/L^6$ van der Waals interaction \cite{Saffman2010,Walker2008}.
In the case $n_{\text{\tiny{A}}}=n_{\text{\tiny{B}}}$, this coupling is dominant and induces a shift of the doubly excited Rydberg state commonly referred to as the blockade shift.
When $n_{\text{\tiny{A}}}\neq n_{\text{\tiny{B}}}$, and under special conditions, the coupling $C^{\mathrm{ex}}_6/L^6$, which exchanges the electronic spin states, may become equally large. In the two-atom product basis $|$$\Uparrow \Downarrow\rangle,~|$$\Downarrow\Uparrow \rangle$, the total van der Waals interaction is then,
\begin{eqnarray}
 {H}_{\text{v}} &=&   \frac{1 }{L^6}\left(\begin{array}{cc} C_6 &  C^{\mathrm{ex}}_6\\
C^{\mathrm{ex}}_6 &C_6 \end{array}
 \right).\label{eq01}
\end{eqnarray}
By using the measured results for the relevant quantum defects~\cite{Li2003} and a semiclassical expression for the radial matrix elements~\cite{Kaulakys1995}, we numerically evaluate $C_6$ and $C^{\mathrm{ex}}_6$ for $0\leq n_{\text{\tiny{B}}}-n_{\text{\tiny{A}}}\leq 10$ and $50<n_{\text{\tiny{B}}}<129$. We find four pairs of states where the interaction coefficients differ by less than $2\%$; see Table~\ref{table1}.

The occurrence of a strong spin-exchange interaction in these cases arises through an interference effect involving a small number of dominant intermediate $p_{1/2}$ and $p_{3/2}$ states~(see Appendix~\ref{appB}). We find that the transition matrix elements for spin exchange constructively interfere whereas a partially destructive interference limits the blockade shift. This results in almost equal magnitude of $C_6$ and $C^{\mathrm{ex}}_6$.

\begin{table}
  \begin{tabular}{|c|c||r|r|}
    \hline     $ n_{\text{\tiny{A}}} $  & $ n_{\text{\tiny{B}}} $ &  $C_6/(2\pi) $ &  $C^{\mathrm{ex}}_6/(2\pi) $ \\ \hline
  59  &	 61  &	 	 -196 &194	 \\
73  &	 75  &		4080  &	 -4025\\
97  &	 100  &		 -59780	  & 58800\\	
121  &	 124  &	 1104000	  & -1124000 \\  \hline
  \end{tabular}
  \caption{  \label{table1} Strength of blockade and spin-exchange interactions between two atoms of principle quantum numbers $n_{\text{\tiny{A}}}$ and $n_{\text{\tiny{B}}}$. $C_6, C^{\mathrm{ex}}_6$ are in unit of $\text{GHz}~\mu \text{m}^6$. }
 \end{table}

The eigenstates and eigenvalues of Eq.~(\ref{eq01}) are $|r_\pm\rangle  \equiv  (|\Uparrow \Downarrow\rangle\pm|\Downarrow\Uparrow \rangle )/\sqrt2$, and $V_\pm \equiv   (C_6\pm C^{\mathrm{ex}}_6)/ L^6$. If the interaction coefficients have equal magnitudes, one of the two energy eigenvalues is unshifted from the non-interacting value.
For the four cases shown in Table~\ref{table1}, $C_6$ and $C^{\mathrm{ex}}_6$ have opposite signs, and by choosing an appropriate atomic separation $L$, $V_-$~($V_+$) can be made much larger~(smaller) than the excitation Rabi frequency. As a result, atoms $A$ and $B$ will be excited to the entangled two-atom Rydberg state $|r_+\rangle$.
The lifetime of the entanglement created can be enhanced by coupling it to the two-atom ground state, by driving Rabi oscillations $|$$\downarrow\rangle\leftrightarrow |$$\Uparrow \rangle$ and $|$$\uparrow\rangle\leftrightarrow  |$$\Downarrow\rangle $ simultaneously on both atoms~[Fig.~\ref{fig001}(b)], so that
the Rydberg triplet state $|r_+\rangle$ is mapped to the ground level Bell state
\begin{equation}
|g_{+}\rangle\equiv (|\uparrow \downarrow\rangle + |\downarrow\uparrow \rangle )/\sqrt2.
\end{equation}

\begin{figure}
\includegraphics[width=3.3in]
{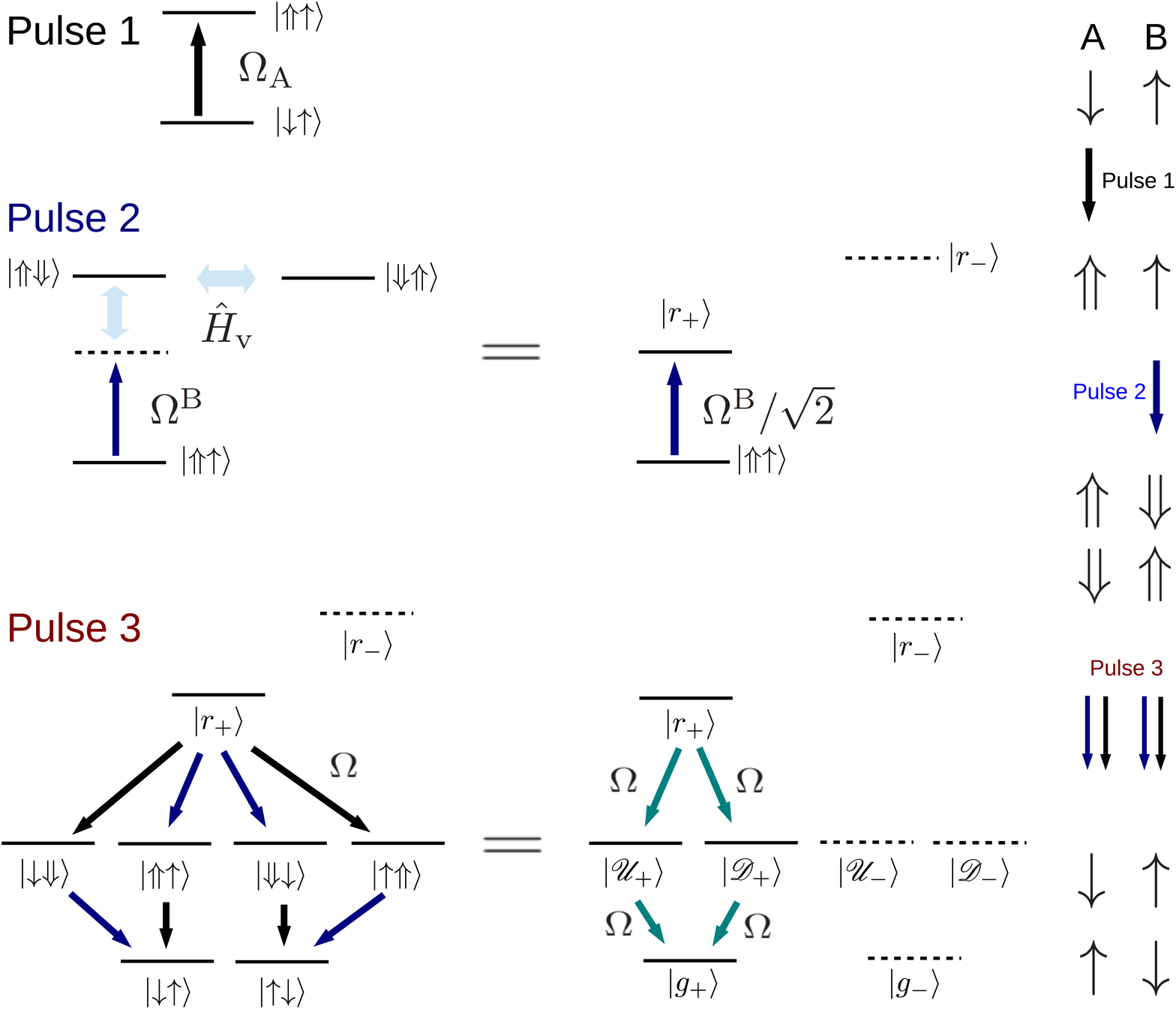}
 \caption{(Color online) Schematic of the entanglement protocol via pulses 1, 2 and 3. During pulse 1~(2), only the two-photon transition via 5P~(6P) state of atom $A$~($B$) is excited with Rabi frequency $\Omega_A$~($\Omega^{B}$). During pulse 3, the four two-photon transitions via 5P and 6P states of atoms $A$ and $B$ are turned on and the magnitudes of these two-photon Rabi frequencies are set equal to $\Omega$. On the left column we show the bare two-atom basis, while on the right we highlight the superposition states involved. \label{fig01a2}}
\end{figure}

\section{Pairwise entanglement protocol} We now describe the complete protocol creating ground state entanglement via a three-$\pi$-pulse sequence. 

{\it Pulse 1 on atom $A$}:
We take the initial state to be $|$$\downarrow\uparrow \rangle$. The first pulse acts on atom $A$ and excites $|$$\downarrow\rangle$ to $|$$\Uparrow \rangle$ via the $5P$ state, as shown in Fig.~\ref{fig01a2}. Since atom $B$ is in its ground state, there is no Rydberg interaction. Thus, by applying a pulse of duration $\tau_1=\pi/\Omega_{\text{A} }$, we generate the product state $|$$\Uparrow \uparrow\rangle$.

{\it Pulse 2 on atom B}: Following pulse 1, apply a two-photon laser pulse to atom $B$ with Rabi frequency $\Omega^{\text{B} }$, as shown in Fig.~\ref{fig01a2}. This pulse excites $|$$\uparrow\rangle$ to $|$$\Downarrow\rangle$ via the $6P$ state as in Fig.~\ref{fig001}(b). The evolution of the two-atom wave function $|\Psi\rangle$ is governed by the Hamiltonian
\begin{eqnarray}
 {H}_{\text{Pulse2}}\!  &\!=\!& \!  \frac{1 }{2\sqrt2}\left( \begin{array}{ccc} 0&\Omega^{\text{B} } & \Omega^{\text{B} }   \\
\Omega^{\text{B} }  &2\sqrt2 V_+ &0\\
\Omega^{\text{B} }   &0&2\sqrt2 V_-
\end{array}
\right), \label{pulse2}
\end{eqnarray}
where the basis vectors are $|$$\Uparrow \uparrow\rangle$,$|r_{+}\rangle$ and $|r_{-}\rangle$. In the case of $|V_-|$$\gg \Omega^{\text{B} } $, we can adiabatically eliminate the state $|r_{-}\rangle$. The population on the Rydberg Bell state $|r_{+}\rangle $ reaches its maximum
\begin{equation}
|\langle r_{+} | \Psi\rangle| \approx 1 -  (V_+/ \Omega^{\text{B} }) ^2,
\end{equation}
for a pulse of duration $\tau_2=(\sqrt2\pi/\Omega^{\text{B}}) \left [ 1 -  (V_+/ \Omega^{\text{B} }) ^2  \right]$.
The prefactor $\sqrt2\pi/\Omega^{\text{B}},$ indicates that this is a two-atom $\pi$-pulse, while the correction $V_+^2/( \Omega^{\text{B} }) ^2$ results from the small shift of $|r_+ \rangle$.  

{\it Pulse 3 on both atoms}: The final step is the mapping of the entangled state of Rydberg atoms onto the ground states. In contrast to Pulses 1 and 2, switch on all four Rabi channels simultaneously exciting the two atoms, as shown in Fig.~\ref{fig001}(b).

When the Rabi frequencies satisfy $\Omega\equiv \Omega_{\text{A} }= \Omega^{\text{A} } = \Omega_{\text{B} } =\Omega^{\text{B} },$ only the intermediate states $|\mathscr{U}_+\rangle$ and $|\mathscr{D}_+\rangle$ are coupled to $|r_{+}\rangle$ and $|g_{+}\rangle,$ where $|\mathscr{U}_\pm\rangle \equiv [  | \Uparrow \uparrow\rangle \pm |\uparrow\Uparrow  \rangle]/\sqrt2$ and $|\mathscr{D}_\pm\rangle \equiv [ | \Downarrow\downarrow \rangle\pm   |\downarrow \Downarrow\rangle]/\sqrt2$, see Fig.~\ref{fig01a2}.
 Ordering the states $|r_{+}\rangle,~|\mathscr{U}_+\rangle,~|\mathscr{D}_+\rangle $ and $|g_{+}\rangle $ the Hamiltonian matrix during Pulse 3 reads,
\begin{eqnarray}
{H}_{\text{Pulse3}}&=&\frac{1}{2}\left( \begin{array}{cccc}2 V_+&\Omega  & \Omega &0 \\
\Omega   & 0&0&\Omega  \\
\Omega   & 0&0&\Omega  \\
0& \Omega    &\Omega   &0
\end{array}
\right). \label{eq05}
\end{eqnarray}
At the end of Pulse 3, of duration $\pi/\Omega$, the entangled Rydberg state $|r_+\rangle$ would be completely mapped onto the entangled ground state $|g_+\rangle$ with fidelity limited by the residual shift $V_+$, and the radiative decay rate of the Rydberg level.

\begin{figure}
\includegraphics[width=3.0in]
{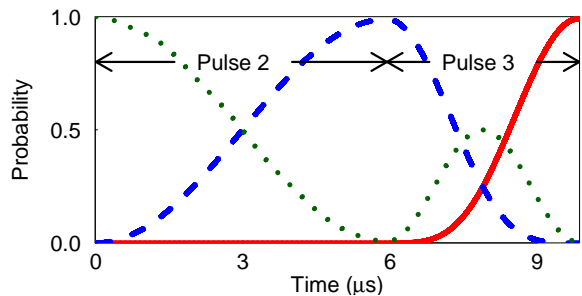}
 \caption{(Color online) Evolution of the two-atom state during pulses 2 and 3 of the entanglement protocol, starting from $|$$\Uparrow \uparrow\rangle$. We plot the populations of the relevant states from the basis in Eq.~(\ref{eq07}): the red (gray) solid line is $|\mathcal{G}_{+}|^2$, the blue (dark gray) dashed line is $|\mathcal{R}_{+}|^2$, the green (light gray) dotted line is $\sum_{\eta}[|$$\mathcal{U}_\eta |^2+ |$$\mathcal{D}_\eta |^2] $. \label{fig003} }
\end{figure}

\section{Numerical simulation and fidelity}
 The proposed scheme relies on creation of the Rydberg entangled state, $|r_{+}\rangle,$ which requires
\begin{equation}
|V_-|\gg \Omega \gg |V_+|.
\end{equation}
In order to show that the ground Bell state $|g_+\rangle$ can be prepared with high fidelity for finite $|V_-/V_+| $, we numerically study the time evolution of the atomic state following the procedure discussed above. We choose the Rydberg state pair with principal quantum numbers $n_{\text{\tiny{A}}}=73,~n_{\text{\tiny{B}}}=75 $, atomic separation $L=15\mu$m, and set the single-atom Rabi frequency $\Omega=\Omega^{\text{B}}=\sqrt{|V_-V_+|} = 2\pi \times 59$ kHz.
We expand the two-atom wavefunction as
\begin{eqnarray}
|\Psi\rangle&=&\sum_{\eta=\pm} \left[\mathcal{ G}_{\eta} |g_{\eta}\rangle+ \mathcal{R}_{ \eta} |r_{\eta}\rangle + \mathcal{U}_{ \eta} |\mathscr{U}_{\eta}\rangle + \mathcal{D}_{ \eta} |\mathscr{D}_{\eta}\rangle \right], \label{eq07}
\end{eqnarray}
and we numerically solve the Schr\"{o}dinger equation for Pulses 2 and 3~(see Appendices~\ref{appC} and~\ref{appD}), since pulse 1 is trivial. The achieved ground state fidelity is $\mathcal{F} =|\langle g_+|\Psi\rangle| ^2=98.5\%$.
To improve the fidelity of the prepared ground Bell state, we numerically optimize $\mathcal{F}$ by varying Rabi frequencies and pulse durations. Figure~\ref{fig003} shows the state evolution with $\Omega^{\text{B}}/(2\pi) = 119$ kHz and $\Omega/(2\pi) = 128$ kHz: a fidelity of $\mathcal{F}_{1}=99.06\%$ for the ground Bell state $|g_+\rangle$ is achieved in less than ten microseconds.

The main practical difficulty is to have all four Rabi frequencies for the excitation channels equal to each other. In order to show that the the protocol is robust against dispersion in the Rabi frequencies, we numerically integrate the Schr\"{o}dinger equation varying $\Omega_{\text{A} }, \Omega^{\text{A} },\Omega_{\text{B} }$ and $\Omega^{\text{B} }$ in the interval $[1-\epsilon,~1+\epsilon]\Omega$ for Pulse 3, with $\Omega=\sqrt{V_+V_-}$, $\tau_2=(\sqrt2\pi/\Omega) \left [ 1 -  (V_+/ \Omega) ^2  \right]$, and $\tau_3=\pi/\Omega$. By performing $10^5$ such simulations, we find that almost all fidelities are larger than $95\%$~($85\%$) for $\epsilon=0.1~(0.2)$~(see Appendix~\ref{appD}). 

\section{SWAP gate and entanglement of atomic chains} 
\begin{figure}
\includegraphics[width=2.20in]
{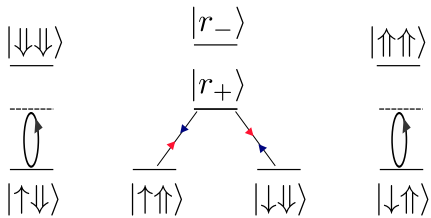}
 \caption{(Color online) Schematic of the SWAP-like gate protocol. During the $2\pi$ pulse,  $|$$r_-\rangle$, $|\Downarrow\Downarrow\rangle$ and $|\Uparrow\Uparrow\rangle$ are hardly excited due to Rydberg blockade, while $|$$r_+\rangle$ will be excited, forming a $\Lambda$ system with $|$$\downarrow\Downarrow\rangle$ and $|$$\uparrow\Uparrow\rangle$. \label{fig005} }
\end{figure}

The spin-exchange Rydberg interaction may be used together with Rydberg blockade to implement a quantum logic gate based on a simple combination of three single-atom laser interaction processes: two $\pi$-pulses applied to atom $B$, are separated by an intermediate $2\pi$-pulse applied to atom A.
If both atoms are initially prepared in the same ground state, we obtain the state transformations,
\begin{eqnarray}
&& |\uparrow \uparrow \rangle \xrightarrow[\text{atom B}]{\pi }
|\uparrow\Downarrow\rangle \xrightarrow[\text{atom A}]{2\pi }  |\uparrow\Downarrow\rangle\xrightarrow[\text{atom B}]{\pi }|\uparrow \uparrow\rangle ,
\nonumber \\
&&| \downarrow\downarrow\rangle\xrightarrow[\text{atom B}]{\pi } |\downarrow\Uparrow\rangle
\xrightarrow[\text{atom A}]{2\pi } |\downarrow\Uparrow\rangle\xrightarrow[\text{atom B}]{\pi }|\downarrow\downarrow \rangle.
\nonumber
\end{eqnarray}
During stages 1 and 3, a single Rydberg excitation is created and removed, while in stage 2, Rydberg blockade of the states $|\Downarrow\Downarrow\rangle$ and $|\Uparrow\Uparrow\rangle$, respectively, prevents any double excitation.
Alternatively, when the atoms are initially prepared in opposite spin ground states, there is a crucial difference. As a consequence of the strong spin-exchange interaction, the 2$\pi$ pulse resonantly couples to the triplet state $| r_+\rangle$ and flips the spin of both the Rydberg excitation and the atomic ground state; see Fig.~\ref{fig005}. In this case the combination of van der Waals interaction and single atom 2$\pi$ pulse creates a resonant ``lambda" transition between two-atom states:
\begin{eqnarray}
&&|\downarrow\uparrow \rangle \xrightarrow[\text{atom B}]{\pi }
 |\downarrow\Downarrow\rangle \xrightarrow[\text{atom A}]{2\pi (\Lambda)}- |\uparrow\Uparrow\rangle\xrightarrow[\text{atom B}]{\pi }-|\uparrow \downarrow\rangle ,
\nonumber \\
&&|\uparrow \downarrow\rangle\xrightarrow[\text{atom B}]{\pi }
 |\uparrow\Uparrow\rangle\xrightarrow[\text{atom A}]{2\pi (\Lambda)}- |\downarrow\Downarrow\rangle \xrightarrow[\text{atom B}]{\pi }-|\downarrow\uparrow \rangle. \nonumber
\end{eqnarray}
The three stage protocol completes the quantum SWAP gate transformation, with a $\pi$ phase shift of the swapped states.
%
By choosing $L=15\mu$m and Rabi frequency $89$kHz for the $2\pi$ pulse on atom A, we can demonstrate a gate fidelity $\mathcal{F}_{\mathrm{SWAP}}=98.31\%$ for an operation time $\tau_{\text{SWAP}} = 11\mu$s.\\

\begin{figure}
\includegraphics[width=3.3in]
{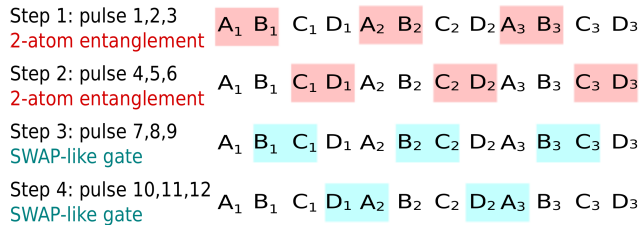}
 \caption{(Color online) Schematic illustration of a protocol to generating entanglement between 12 $^{87}$Rb atoms loaded in a one-dimensional lattice. The pairwise entanglement protocol creates entanglement between atoms $A_j$ and $B_j$~($C_j$ and $D_j$) during step 1~(2), while the SWAP-like gate protocol entangles atoms $B_j$ and $C_j$~($D_j$ and $A_{j+1}$) during step 3~(4).  \label{figIllus}   }
\end{figure}

The combination of quantum SWAP gate with pairwise atom entanglement operations can be used as the basis of a protocol to entangle a chain of an arbitrary number of atoms.
The key observation is that the atoms should be entangled sequentially in pairs, allowing gaps between the pairs in order to minimize spurious level shifts due to Rydberg blockade. After the pair-entanglement, a series of SWAP gate operations link all the pairs in a fully entangled state.
Such a procedure may be easily sketched for the case of $4N$ atoms, where the atoms are labeled sequentially: $A_1,~B_1,~C_1,~D_1,~\cdots,A_{4N},~B_{4N},~C_{4N},~D_{4N}$.
We first use the two-atom entanglement protocol described above to entangle atoms $A_j$ and $B_j$, where $j=1,\cdots,N$. This is followed by a similar protocol that entangles atoms $C_j$ and $D_j$. Because all atoms are now entangled with one of their two neighboring atoms, the two-qubit SWAP gate is used to entangle atoms $B_j$ and $C_j$, followed by another SWAP operation for entanglement of $D_j$ and $A_{j+1}$. In this way we may entangle all $4N$ atoms by a twelve-pulse protocol as shown in Fig.~\ref{figIllus}.
In the case of 4 atoms, this protocol generates the entangled state $\left( |\uparrow \downarrow\downarrow\uparrow\rangle+|\downarrow\uparrow\uparrow \downarrow \rangle  - |\uparrow   \uparrow\downarrow  \downarrow\rangle-  |\downarrow \downarrow \uparrow\uparrow\rangle\right)/2, $ after the application of 9 pulses~(see Appendix~\ref{appE}).

During the pairwise entanglement and SWAP gate operations, the metastable Rydberg states are populated for a time $\tau$, thus the fidelity of the entanglement of the $4N$-atom chain may be estimated as $\mathcal{F}\sim(\mathcal{F}_{1}e^{-2\gamma \tau })^{2N}(\mathcal{F}_{\mathrm{SWAP}}e^{-2\gamma \tau})^{2N-1}$, where $\gamma^{-1}$ is the lifetime of the Rydberg level. When keeping the leading order term after expanding the exponential of $\mathcal{F}$ under the condition of $\gamma \tau\ll1$, the total error of the prepared many-atom entangled state scales linearly with $N$. 
Numerical simulations show that pairwise entanglement and SWAP gate operations can each be carried out in around 10$\mu$s. Thus a 4-atom chain may be entangled in a time $T= 30\mu$s with $L=15\mu$m. For (a) $(n_A,n_B) = (73,75)$, $\gamma^{-1} = 0.45$~ms for $n=73$ and $\mathcal{F}\sim90\%$, while for (b) $(n_A,n_B) = (97,100)$, $\gamma^{-1} = 1$~ms for $n=97$ and $\mathcal{F}\sim 94\%$ ~\cite{Saffman2010}. Decreasing $L$ to $11.9\mu m$ makes $V_\pm$ four times larger, and reduces $T$ to $7.5\mu s$. Then For $(n_A,n_B) = (73,75)$, a 4-atom chain can be entangled with $\mathcal{F}\sim95\%$, while an 8-atom chain can be entangled with $\mathcal{F}\sim90\%$. 
These fidelities are comparable to the values for 4 and 8 atom entanglement by asymmetric blockade in Ref.~\cite{PhysRevLett.102.240502} and dissipation in Ref.~\cite{Carr2013}. The dissipative protocol in \cite{Carr2013} does not suffer from the spontaneous emission issue, but in comparison the present coherent process is almost three orders of magnitude faster. Furthermore, the linear scaling of the error with the total number of entangled atoms is similar to the the blockade-based situation of Ref.~\cite{PhysRevLett.102.240502}: in that case the error scales cubically at low $N$ and then saturates to a linear behavior.
By contrast to the multi-atom entanglement based on Ryberg interactions and adiabatic passage proposed in Ref.~\cite{PhysRevLett.100.170504}, the duration and the Rabi frequency of our scheme are independent of $N$.
The requirements on the Rabi frequencies and principal quantum numbers are well within experimental reach and the individual atomic addressing allows to tailor the distance between the atoms as well as the desired target state. 
All these considerations show that the proposed spin-exchange mechanism represents a valid candidate to realize fast quantum operations with Rydberg atoms.

\section{Conclusion}
Throughout this work we have considered the case of individually trapped atoms addressed by external lasers.  An interesting alternative realization of these protocols relies on the creation of a small \emph{super-atom} in an elongated ensemble. This option has the advantage of solving the probabilistic loading of the individual traps as well as to provide a boost to the single-atom procedure by a factor $\sqrt{N}$ where $N$ is the number of individual atoms in the super-atom. Quantum gates between super-atoms have been recently proposed \cite{Paredes-Barato2014} and the multiplexing of different qubits in an elongated ensemble has already been realized experimentally \cite{Alexmulti}. 

In conclusion, we have proposed a fast and robust mechanism to entangle neutral atoms. It is based on a variation of the van der Waals interaction between atoms excited to Rydberg states: for different principal quantum numbers, the spin-exchange interaction may be comparable to the Rydberg blockade shift thus induce a resonance between ground state levels and an entangled Rydberg state. This metastable state may then be mapped to a stable, entangled ground state. Furthermore, the entanglement efficiency may be improved by using small ensembles as well as by manipulating the structure of the Rydberg manifolds via external fields.
Based on the spin-exchange interaction, pairwise entanglement along with a SWAP-like gate form the basis of a protocol for the generation of ground-state entanglement of many atoms in a chain configuration. These protocols may be implemented in present experiments leading to quantum manipulation of many-body systems.\\

\section*{Acknowledgments}
XFS and TABK acknowledge support from AFOSR and the Quantum Memories MURI of the Air Force Office of Scientific Research. FB acknowledges support from the DARPA QuASAR program and the US NSF.

\appendix{}
\section{Single-atom optical excitation}\label{appA}
\begin{figure}
\includegraphics[width=3.3in]
{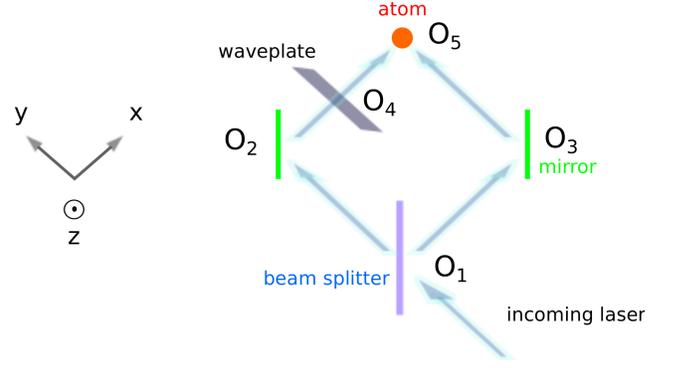}
 \caption{(Color online) Schematic illustration for generating an effective right-hand or left-hand polarized light field from two linearly polarized light fields related by a $\pi/2$ or $3\pi/2$ phase difference. \label{fig006}}
\end{figure}

This section shows how to realize the transitions in Fig~1(b) of the main text.
We consider an atom optically excited via two linearly polarized light fields, one $\mathbf{x}$-polarized, the other $\mathbf{y}$-polarized, traveling along $\mathbf{y}$ and $\mathbf{x}$ direction, respectively. To have such a pair of light fields, a beam splitter is placed along the $\mathbf{z}$ axis as shown in Fig.~\ref{fig006}. For each incoming light field, the following optical devices are used, ($O_1$), beam splitter, ($O_2$) and ($O_3$), mirror, and ($O_4$), a quarter-wave plate, or a wave plate whose thickness makes it effectively a combination of a half wave-plate and a quarter-wave plate, depending on the specific low-lying intermediate P state.
By tuning the positions of the two mirrors so that they are symmetric to each other about the plane of the beam splitter, we have
\begin{eqnarray}
l&\equiv &l_{12}+l_{25} = l_{13}+l_{35}, \label{condlength}
\end{eqnarray}
where $l_{\alpha\beta}$ is the distance between $O_\alpha$ and $O_\beta$, where $\alpha~(\beta)=1,\cdots,5$. Here $O_5$ denotes the position of the atom.
Assuming that the light field impinging on the beam splitter is $\mathbf{x}\mathscr{E}\cos(\omega t)$, one can show that the electric field on the atom is:

\begin{eqnarray}
\mathbf{E}&=& -\frac{\mathscr{E}}{2\sqrt2} \left[ \left( \mathbf{x} + \mathbf{y}e^{i\theta}\right) e^{ i\omega t - il  k } +\text{h.c.}\right ].
\label{field01}
\end{eqnarray}
In order to excite the the Rydberg states $|$$\Uparrow\rangle$ and $|$$\Downarrow\rangle$ defined in the main text for either atom $A$ or atom $B$, one shall use waveplates so that $\theta = \pi/2$ and $3\pi/2$, respectively. 
As a result, the transition from the ground level $|$$\uparrow(\downarrow)\rangle\equiv |5^2S_{\frac{1}{2}}, F = 2, m_F = 2(-1)\rangle$ to the Rydberg level $|$$\Downarrow(\Uparrow)\rangle\equiv |n~  ^2S_{\frac{1}{2}}, m_J = \mp 1/2, m_I =1/2  \rangle$ via an intermediate level $ |6^2S_{\frac{1}{2}}, F = 1, m_F = 1\rangle$~($ |5^2S_{\frac{1}{2}}, F = 1, m_F = 0\rangle$) can be realized by a two-photon Rabi process via a pair of effective left~(right)-hand polarized light fields.

The non-interacting Hamiltonian for atom A~(the Hamiltonian for atom B is similar) in the dipole approximation and rotating-wave approximation for the atom-field coupling is
\begin{eqnarray}
H_{\text{A}}&=& \sum_{m_F} E_g |5S_{1/2},F=2,m_F\rangle\langle 5S_{1/2},F=2,m_F|\nonumber\\
&&
 + \sum_{m_f} E_{5e} |5P_{1/2},F'=1,m_f\rangle\langle 5P_{1/2},F'=1,m_f|\nonumber\\
&&
+ \sum_{m_f} E_{6e} |6P_{1/2},F'=1,m_f\rangle\langle 6P_{1/2},F'=1,m_f|\nonumber\\
&&
+ \big[\sum_{m_J,m_I} E_r |m_J,m_I\rangle\langle m_J,m_I|\nonumber\\
&&+ \sum_{m_F}\sum_{m_f}\Omega_{ 5m_f, m_F}^{(ge)}(q)e^{-i\omega_5^{(ge)}t}\nonumber\\
&&~~~  | 5P_{1/2},F'=1,m_f \rangle\langle 5S_{1/2}, F=2,m_F|  \nonumber\\
&&+\sum_{m_f}\sum_{ m_J,m_I} \Omega_{5m_f,m_Jm_I}^{(er)}(q) e^{i\omega_5^{(er)}t}\nonumber\\
&&~~~ |  5P_{1/2},F'=1,m_f \rangle\langle m_J,m_I|\nonumber\\
&&+ \sum_{m_F}\sum_{m_f}\Omega_{ 6m_f, m_F}^{(ge)}(\overline{q}) e^{-i\omega_6^{(ge)}t}  \nonumber\\
&&~~~ | 6P_{1/2},F'=1,m_f \rangle\langle 5S_{1/2}, F=2,m_F|  \nonumber\\
&&+\sum_{m_f}\sum_{ m_J,m_I}\Omega_{6m_f,m_Jm_I}^{(er)}(\overline{q}) e^{i\omega_6^{(er)}t}\nonumber\\
&&~~~ |  6P_{1/2},F'=1,m_f\rangle\langle m_J,m_I| + \text{h.c.}\big],
\end{eqnarray}
where $q=1 (-1)$ indicates right~(left)-hand polarization, $E_{j}$ with $j =  [g,5e,6e,r]$ is the energy of a specific atomic manifold, $\omega_{5(6)}^{(ge)}$ and $\omega_{5(6)}^{(er)}$ are the central frequencies of the lasers for the lower and upper transitions, and they satisfy
\begin{eqnarray} 
\omega_5^{(ge)} + \omega_5^{(er)} &=& \omega_6^{(ge)} + \omega_6^{(er)}.
\end{eqnarray}
The Rabi frequencies read,
\begin{eqnarray} 
\Omega_{ nm_f, m_F}^{(ge)}(q) &=& \frac{1}{2}D_n^{(ge)}\mathcal{E}_{ge} C_{m_F ~ q ~ m_f}^{F~1~ F'},\nonumber\\ 
\Omega_{nm_f,m_Jm_I}^{(er)}(\overline{q}) &=& \frac{1}{2}D_n^{(er)}\mathcal{E}_{er}\sum_\alpha C_{m_J ~ \overline{q} ~ \alpha}^{\frac{1}{2}~1~ \frac{1}{2}} C_{ \alpha~ m_I ~ m_f}^{\frac{1}{2}~I~ F'}, \nonumber 
\end{eqnarray}
where $n=5$ or 6, $\mathcal{E}_{ge(er)}$ is the electric field of the laser for the lower~(upper) transition in Eq.~(\ref{field01}), $D_n^{(ge)}=e(  nP_{1/2},F=1|| r ||5S_{1/2},F=2)$ and $D_n^{(er)}=e(  nP_{1/2} || r ||n_{\text{A}} S_{1/2})$ are the reduced matrix elements of the electric dipole operator obtained via the Wigner-Eckart theorem~\cite{Rose1957,Jenkins2012}, with $e$ the elementary charge, and $C$ a Clebsch-Gordan coefficient~\cite{Rose1957}. 
\begin{eqnarray}
\frac{\Omega_n^{(ge)}(1)}{  D_n^{(ge)}\mathcal{E}_{ge} } &=&- \left(\begin{array}{ccccc}
	0.775&	0&	0&	0&	0\\	
	0&	0.548&	0&	0&	0\\	
	0&	0&	0.316&	0&	0
\end{array}
\right),\nonumber\\
\frac{\Omega_n^{(ge)}(-1)}{  D_n^{(ge)}\mathcal{E}_{ge} }  &=& \left(\begin{array}{ccccc}
	0&	0&	0.316&	0&	0\\		
	0&	0&	0&	0.548&	0\\		
	0&	0&	0&	0&	0.775
\end{array}
\right),\nonumber\\
\frac{\Omega_n^{(er)}(1)}{  D_n^{(er)}\mathcal{E}_{er} }  &=& -\left(\begin{array}{cccccccc}
	0&	0&	0&	0&	0&	0.408&	0&	0\\	
	0&	0&	0&	0&	0&	0&	0.577&	0\\	
	0&	0&	0&	0&	0&	0&	0&	0.707		
\end{array}
\right),\nonumber\\
\frac{\Omega_n^{(er)}(-1)}{ D_n^{(er)}\mathcal{E}_{er} } &=& \left(\begin{array}{cccccccc}
	0.707&	0&	0&	0&	0&0	&	0&	0\\	
	0&	0.577&	0&	0&	0&	0&0	&	0\\	
	0&	0&	0.408&	0&	0&	0&	0&0		
\end{array}
\right).\nonumber
 \end{eqnarray}
Here, the column and row indices for $\Omega^{(ge)}$ are for quantum numbers $m_F= [-2,-1,0,1,2]$~(from left to right) and $m_f= [-1,0,1]$~(from top to bottom), and the column and row indices for $\Omega^{(er)}$ are for quantum numbers $(m_Jm_I)=[ (\frac{-1}{2}\frac{-3}{2}),(\frac{-1}{2}\frac{-1}{2}),(\frac{-1}{2}\frac{1}{2}),(\frac{-1}{2}\frac{3}{2})$, $(\frac{1}{2}\frac{-3}{2}),(\frac{1}{2}\frac{-1}{2}),(\frac{1}{2}\frac{1}{2}),(\frac{1}{2}\frac{3}{2})]$(from left to right) and $m_f=[-1,0,1]$(from top to bottom).

Below, we derive the effective two-photon Rabi frequency when the two-photon detuning is zero~\cite{James2000,Brion2007,Han2013}. Using the method of Ref.~\cite{James2000}, we can derive a Hamiltonian for far off resonant optical driving. The method of Ref.~\cite{James2000} is essentially an adiabatic approximation. The Hamiltonian in a rotating frame and rotating-wave approximation for a three level system with basis $|r\rangle,~|e\rangle,~|g\rangle$ is~(the subscripts d and u denote down and up, respectively)
\begin{eqnarray}
H_{\text{3level}}&=& \left( \begin{array}{ccc} 0 &e^{ik_{\text{u}}l_{\text{u}}} \Omega_{\text{u}}/2 & 0 \\
e^{-ik_{\text{u}}l_{\text{u}}} \Omega_{\text{u}}/2 & -\Delta &e^{ik_{\text{d}}l_{\text{d}}}  \Omega_{\text{d}}/2 \\
0 & e^{-ik_{\text{d}}l_{\text{d}}} \Omega_{\text{d}}/2 &0\end{array}\right), \nonumber
\end{eqnarray}
where $l_{\text{u(d)}}$ is the length defined in Eq.~(\ref{condlength}) for the transition from $|g\rangle$ to $|e\rangle$~(from $|e\rangle$ to $|r\rangle$). Here we assume that $\Omega_{\text{u(d)}}$ is real, and the two-photon transition is resonant. We write the wave function as $ \alpha_r|r\rangle + \alpha_e  |e\rangle +\alpha_g |g\rangle$, and starting from state $(\alpha_r(0), \alpha_g(0) ) = (0, 1)$, one can find $|\alpha_r(t) |\leq \frac{|\Omega _{\text{d}} \Omega _{\text{u}} | }{\Omega _{\text{d}}^2+\Omega _{\text{u}}^2}\leq 1$ and
\begin{eqnarray}
|\alpha_g(t) |^2 &=& 1+  \frac{2 \Omega _{\text{d}}^2\Omega _{\text{u}}^2  }{(\Omega _{\text{d}}^2+\Omega _{\text{u}}^2)^2} \left[\cos\frac{ t \left(\Omega _{\text{d}}^2+\Omega _{\text{u}}^2\right)}{4 \Delta } -1\right],\nonumber
\end{eqnarray}
which indicates that Max$|\alpha_g(t) |^2$ and Min$|\alpha_g(t) |^2$ achieve maximum difference only when $|\Omega _{\text{d}}|= |\Omega _{\text{u}}|$. This can guide setting up the condition in experiments. In fact, when the difference of Max$|\alpha_g(t) |^2$ and Min$|\alpha_g(t) |^2$ reaches its maximum while adjusting one of $|\Omega _{\text{d}}|, |\Omega _{\text{u}}|$, the condition $|\Omega _{\text{d}}|= |\Omega _{\text{u}}|$ is met. When $\Omega_{\text{d}}=\Omega_{\text{u}}$, the Hamiltonian in a rotating frame is
\begin{eqnarray}
H&=&\frac{\Omega_{\text{d}}\Omega_{\text{u}}}{4\Delta} \left( \begin{array}{cc}  0 &e^{i(k_{\text{d}}l_{\text{d}}+k_{\text{u}}l_{\text{u}})}  \\
e^{-i(k_{\text{d}}l_{\text{d}}+k_{\text{u}}l_{\text{u}})} & 0\end{array}\right). \nonumber
\end{eqnarray}
When the lower and upper Rabi transitions have the same Rabi frequency, the effective Rabi frequency between $|r\rangle$ and $|g\rangle$ is $\frac{\Omega_{\text{d}}\Omega_{\text{u}}}{2\Delta}  e^{i(k_{\text{d}}l_{\text{d}}+k_{\text{u}}l_{\text{u}})}$. By assuming $\frac{\Omega_{\text{d}}\Omega_{\text{u}}}{2\Delta}>0$, we can write the effective Rabi frequency as $|\frac{\Omega_{\text{d}}\Omega_{\text{u}}}{2\Delta}| e^{i(k_{\text{d}}l_{\text{d}}+k_{\text{u}}l_{\text{u}})}$.

For the system studied in the main text, we shall identify $k_{\text{d}}l_{\text{d}}+k_{\text{u}}l_{\text{u}}$ as $-\phi_{\text{A} }$ for the two-photon transition via 5P state of atom $A$, while as $-\phi^{\text{A} }$ for the two-photon transition via 6P state of atom $A$. But since both of these two-photon transitions are resonant, and the ground states or the Rydberg states are degenerate, we immediately find $\phi_{\alpha }= \phi^{\alpha}$ if one sets a common $l\equiv l_{\text{d}}=l_{\text{u}}$ for exciting both of the two Rydberg state $|$$\Uparrow\rangle_\alpha$ and $|$$\Downarrow\rangle_\alpha$, with $\alpha=A$ or $B$. 

\section{van der Waals interaction}\label{appB}
Since we consider the uncommon situation of the interaction between two Rydberg levels of different principal quantum numbers, we will briefly outline a perturbation calculation here. Consider two $^{87}$Rb Rydberg atoms, one prepared in state $|r_{1\pm}\rangle = |n_1~^2S_{\frac{1}{2}}, m_J = \pm 1/2\rangle $, and the other in state $|r_{2\pm}\rangle = |n_2~^2S_{\frac{1}{2}}, m_J = \pm 1/2 \rangle $, where $n_1\neq n_2$. We consider the following four channels for the dipole-dipole interaction, each characterized by its energy defect~\cite{Walker2008},
\begin{eqnarray}
\delta_1(n_s,n_t)&=& E(n_sp_{\frac{3}{2}}) + E(n_tp_{\frac{3}{2}}) -  E(n_1 s_{\frac{1}{2}})-  E(n_2 s_{\frac{1}{2}}),\nonumber\\
\delta_2(n_s,n_t)&=& E(n_sp_{\frac{3}{2}}) + E(n_tp_{\frac{1}{2}}) -  E(n_1s_{\frac{1}{2}})-  E(n_2 s_{\frac{1}{2}}),\nonumber\\
\delta_3(n_s,n_t)&=& E(n_sp_{\frac{1}{2}}) + E(n_tp_{\frac{3}{2}}) -  E(n_1 s_{\frac{1}{2}})-  E(n_2 s_{\frac{1}{2}}),\nonumber\\
\delta_4(n_s,n_t)&=& E(n_sp_{\frac{1}{2}}) + E(n_tp_{\frac{1}{2}}) -  E(n_1 s_{\frac{1}{2}})-  E(n_2 s_{\frac{1}{2}}).\nonumber\\
\label{delta4}
 \end{eqnarray}
Here $(n_s,n_t)$ denote the principal quantum numbers of the pair state produced by the scattering process. These four couplings are known to be the dominant ones in our case. In the van der Waals interaction the atoms then go back to the initial levels and the magnetic quantum number of either atom can change up to $1$ unit \cite{Walker2008}. 
We can separate the angular dependence of the interaction from the principal quantum numbers. Its matrix representation in the basis of $|r_{1+} r_{2+}\rangle,(|r_{1-} r_{2+}\rangle + |r_{1+} r_{2-}\rangle)/\sqrt2,(|r_{1-} r_{2+}\rangle - |r_{1+} r_{2-}\rangle)/\sqrt2,|r_{1-} r_{2-}\rangle $ is given by
\begin{eqnarray}
D_1 &=&\text{diag}(22,34,18,22)/81,\nonumber\\ 
D_2 &=&  D_3 =\text{diag}(14,2,18,14)/81,\nonumber\\
D_4 &= &\text{diag}(4,16,0,4)/81 
,\label{van001}
 \end{eqnarray}
where diag$(a_1,a_2,a_3,a_4)$ means a diagonal matrix with $a_i$ as diagonal matrix elements. In the basis of $|r_{1+} r_{2+}\rangle$,$|r_{1-} r_{2+}\rangle$, $|r_{1+} r_{2-}\rangle$, and $|r_{1-} r_{2-}\rangle $, Eq.~(\ref{van001}) becomes,

\begin{eqnarray}
D(\delta_1)&=&  \frac{1}{81}\left(\begin{array}{cccc} 22&0&0&0\\
0& 26& 8 &0\\
0& 8 & 26&0\\
0&0&0&22\end{array}
\right),\nonumber\\
D(\delta_{2(3)}) &=&   \frac{1}{81}\left(\begin{array}{cccc} 14&0&0&0\\
0& 10& -8 &0\\
0& -8 & 10&0\\
0&0&0&14\end{array}
\right),\nonumber\\
D(\delta_4) &=&  \frac{1}{81} \left(\begin{array}{cccc} 4&0&0&0\\
0& 8& 8 &0\\
0& 8 & 8&0\\
0&0&0&4\end{array}
\right).\nonumber
\end{eqnarray}

The van der Waals interaction strength for each channel is
\begin{eqnarray}
C_{6}^{(k)} &= &\sum_{\Delta n_{\text{A}},\Delta n_{\text{B}}  = -\Delta n}^{\Delta n} \mathcal{C}_6 (\delta_k(n_{\text{A}}+\Delta n_{\text{A}},n_{\text{B}}+\Delta n_{\text{B}}  )).\nonumber
 \end{eqnarray}
We can identify two types of $\mathcal{C}_6$ coefficients:
\begin{eqnarray}
&&\mathcal{C}_6 (\delta_k  ) = \frac{-e^4}{\delta_i(n_{\text{A}}+\Delta n_{\text{A}},n_{\text{B}}+\Delta n_{\text{B}}  )}\nonumber\\
&&\left( \int rP_{n_{\text{A}} ,l_{\text{A}}} P_{n_{\text{A}} +\Delta n_{\text{A}}, l_{\text{A}}'}dr  \int rP_{n_{\text{B}} , l_{\text{B}}} P_{n_{\text{B}} +\Delta n_{\text{B}},  l_{\text{B}}'}dr\right)^2, \nonumber\\
&&\mathcal{C}'_6(\delta_k) =\frac{-e^4}{\delta_i(n_{\text{A}}+\Delta n_{\text{A}},n_{\text{B}}+\Delta n_{\text{B}}  )}\nonumber\\ &&
\int rP_{n_{\text{A}} ,l_{\text{A}}} P_{n_{\text{A}} +\Delta n_{\text{A}}, l_{\text{A}}'}dr  \int rP_{n_{\text{B}} , l_{\text{B}}} P_{n_{\text{B}} +\Delta n_{\text{B}},  l_{\text{B}}'}dr\nonumber\\ &&
\int rP_{n_{\text{B}} , l_{\text{B}}} P_{n_{\text{A}} +\Delta n_{\text{A}}, l_{\text{A}}'}dr  \int rP_{n_{\text{A}} ,l_{\text{A}}} P_{n_{\text{B}} +\Delta n_{\text{B}},  l_{\text{B}}'}dr, 
 \end{eqnarray}
with $i=1,2,3$ or $4$. Here $P_{n_{1(2)} ,l_{1(2)}}$ is the radial part of the atomic wave function, and the integration about $P_{n_{1(2)} ,l_{1(2)}}$ can be approximated as in Ref.~\cite{Kaulakys1995}.
\begin{table}
  \begin{tabular}{|l||c|c|c|}
    \hline   Value of $\Delta n$  & $ {C}_{6}^{(2)}~({C}_{6}^{(3)})$   \\ \hline
  1 	&	71841.8	\\
  2 	&	71922.5	 \\
  3 		&71928.5	 \\
  6 		 & 	71929.9	 \\
  10 	 & 	71930 	  \\
  15 	 & 	71930 	  \\
  20 	 & 	71930 	 \\  \hline
  \end{tabular}
  \caption{  \label{table2}  Convergence of the van der Walls interaction strength depending on the channels considered (up to $\Delta n$) for a pair of atoms in the state $|100^2S_{\frac{1}{2}} \rangle $. Unit: $\text{GHz}\mu \text{m}^6$.}
\end{table}
When $n_1=n_2$, we have $\mathcal{C}_6 = \mathcal{C}'_6$. Also, the two channels with energy defect $\delta_2$ and $\delta_3$ are the same. We report in Table~\ref{table2} the values of $C_{6}^{(2)}$ or $C_{6}^{(3)}$ for different $\Delta n$ for two atoms at $n_1=n_2=100, l_1=l_2=0$.  The total van der Waals interaction $\sum_xC_{6}^{(x)}D_x/(2\pi)$ is
 \begin{eqnarray}
\hat{H}_{\text{vdW}}^{(100,100)} \!\! &\!\!=\!\!&\!\!  \left(\begin{array}{cccc}56200 &0&0&0\\
0&56980 & 1573 &0\\
0& 1573 &56980 &0\\
0&0&0&56200 \end{array}
\right)\frac{\mu \text{m}^6 \text{GHz}}{L^6},\nonumber\\
\end{eqnarray}
where $L$ is the distance between the two atoms.
The van der Waals interaction for two atoms in the $n_1=97,n_2=100, l_1=l_2=0$ levels shall be calculated considering that the two atoms may swap their principal quantum numbers, i.e., we shall take into account the two cases with $\mathcal{C}_6$ and $\mathcal{C}'_6$. The calculation gives us 
\begin{eqnarray}
\hat{H}_{\text{vdW}}^{(97,100)} &=& \left(\begin{array}{cc}
\hat V_1 &\hat  V_2\\
\hat V_2 &\hat  V_1\end{array}
 \right),
 \end{eqnarray}
where 
\begin{eqnarray}
\hat V_1&=& -\left(\begin{array}{cccccccc}
 89180 &0&0&0\\
 0&59780	  & 	- 58800 &0  \\
0&- 58800 & 59780 &0 \\
0 &0 &0 &  89180
\end{array}
 \right)\frac{\mu \text{m}^6 \text{GHz}}{L^6} ,\nonumber\\
\hat V_2&=&\left(\begin{array}{cccccccc}
-537  &0 &0 &0\\
0 &-375 & 324 &0 \\
0& 324 &-375 &0 \\
0 &0 &0&-537
\end{array}
 \right)\frac{\mu \text{m}^6 \text{GHz}}{L^6}, \label{eq101503}
 \end{eqnarray}
and the basis vectors of $\hat{H}_{\text{vdW}}^{(97,100)}$ for the columns from left to right, and for the rows from top to bottom are
\begin{eqnarray}
&&| r_{\text{A}+}  r_{\text{B}+}\rangle,| r_{\text{A}-}  r_{\text{B}+}\rangle , | r_{\text{A}+}  r_{\text{B}-}\rangle,| r_{\text{A}-}  r_{\text{B}-}\rangle,| r_{\text{B}+}  r_{\text{A}+}\rangle,\nonumber\\
&&~~| r_{\text{B}-}  r_{\text{A}+}\rangle , | r_{\text{B}+}  r_{\text{A}-}\rangle,| r_{\text{B}-}  r_{\text{A}-}\rangle.
\end{eqnarray}
From this matrix, we find that the transition rate to a pair state with exchanged quantum numbers is only $0.6\%$ of that to a state pair with the same quantum numbers. Thus, we can safely ignore these terms. In this approximation, the van der Waals interaction is, in the basis of $|r_{1+} r_{2+}\rangle,|r_{1-} r_{2+}\rangle , |r_{1+} r_{2-}\rangle,|r_{1-} r_{2-}\rangle $
\begin{eqnarray}
\hat{H}_{\text{vdW}}^{(97,100)} &\approx& \left(\begin{array}{cc}
\hat V_1 &0\\
0 &\hat  V_1\end{array}
 \right).
 \end{eqnarray}
 In the two-atom product basis $|$$ r_{\text{A}-}  r_{\text{B}+}\rangle,~|$$r_{\text{A}+}  r_{\text{B}-} \rangle$, the total van der Waals interaction is then,
\begin{eqnarray}
 \hat{H}_{\text{v}} &=&   \frac{1 }{L^6}\left(\begin{array}{cc} C_6 &  C^{\mathrm{ex}}_6\\
C^{\mathrm{ex}}_6 &C_6 \end{array}
 \right), \label{eqB9}
\end{eqnarray}
where the four matrix elements are given in $\hat V_1$ of Eq.~(\ref{eq101503}) for the case of $n_{\text{A}}=97,n_{\text{B}}=100$. This implies that $C_6/(2\pi) = -59780~[\text{GHz}~\mu \text{m}^6]$ and $C_6^{\text{ex}}/(2\pi)=  58800~ [\text{GHz}~\mu \text{m}^6]$ in the main text.

\subsection{Critical radius for van der Waals interaction}
Since we derive the van der Waals strength via a perturbative calculation based on the dipole-dipole potential, it is useful to study the transition from the dipole-dipole interaction to the van der Waals interaction. We consider the relevant case of $n_1=97,n_2=100, l_1=l_2=0$ again. From Ref.~\cite{Walker2008}, the dipole-dipole interaction can be written as
\begin{eqnarray}
V_{n_1,n_2}^{\text{dd}} &=& R_{\text{rr}} M_{j_1'm_1'j_2'm_2';j_1m_1j_2m_2} /\mathcal{L}^3 ,
 \end{eqnarray}
where $M$ is a matrix characterizing the angular dependence of the interaction~\cite{Walker2008}, and
\begin{eqnarray}
R_{\text{rr}} &=& e^2 \int rP_{n_1 ,l_1} P_{n_1 +\Delta n_1, l_1'}dr  \int rP_{n_2 ,l_2} P_{n_2 +\Delta n_2, l_2'}dr. \nonumber
 \end{eqnarray}
The coefficient $R_{\text{rr}}$ is closely related to the usual $C_3$ matrix element.
In Table~\ref{table3}, all the two-atom Rydberg p-states with an energy defect $\delta_e$ smaller than $1$~GHz with respect to the state $n_1=97,n_2=100, l_1=l_2=0$, and $R_{\text{rr}}$ larger than $1$~GHz$\mu m^3$ are listed, from which one finds that the condition $\Delta n_1=0~(2), \Delta n_2 = -1(-3)$ gives the smallest energy defects. Since for $\Delta n_1=2, \Delta n_2 = -3$, the radial matrix elements are small, we consider the case $\Delta n_1=0, \Delta n_2 = -1$, where we find

\begin{align}
&M(\delta_1) = \left( \begin{array}{cccc}
0     &0     &0     &0 \\
0     &0     &0     &0 \\
-0.19&0     &0     &0 \\
0     &-0.33&0     &0 \\
0     &0     &0     &0 \\
-0.44&0     &0     &0 \\
0     &-0.44&-0.11&0 \\
0     &0     &0     &-0.19 \\	
-0.19&0     &0     &0 \\
0     &-0.11&-0.44&0 \\
0     &0     &0     &-0.44 \\	
0     &0     &0     &0 \\
0     &0     &-0.33&0 \\
0     &	0     &0     &-0.19 \\	
0     &0     &0     &0 \\
0     &0     &0     &0 
\end{array}
\right),\\
&M(\delta_{2(3)}) = \left( \begin{array}{cccc}
0   &0   &0   &0    \\	
-0.27&0   &0   &0  \\  	
-0.31&0   &0   &0  \\  	
0   &0.31&-0.16&0   \\ 	
0   &0.16&-0.31&0   \\ 	
0   &0   &0   &0.31 \\	
0   &0   &0   &0.27 \\	
0   &0   &0   &0   		
\end{array}
\right), \\
&M(\delta_4) = \left( \begin{array}{cccc}
-0.22&0     &0     &0 \\
0     &0.22&0.22&0 \\
0     &0.22&0.22&0 \\
0     &0     &0     &-0.22	
\end{array}
\right), 
 \end{align}
where the column indices for the three matrices are $(m_1,m_2) = [(\frac{-1}{2},\frac{-1}{2} ),(\frac{-1}{2},\frac{1}{2} ), (\frac{1}{2},\frac{-1}{2} ),(\frac{1}{2},\frac{1}{2} )]$ from left to right, and the row indices are $(m_1',m_2') = [(-j_1',-j_2' ), (-j_1',-j_2'+1 ),\cdots, (j_1',j_2'-1 ) , (j_1',j_2' )] $ from top to bottom. From the three matrices and Table~\ref{table3}, we find that the state (97$p_{1/2}$,~99$p_{3/2})$ is the dominant scattering channel and we can thus determine the critical distance $L_{\text{dd-vdW}}$ where dipole-dipole interaction transitions into the  van der Waals regime, 
\begin{eqnarray}
\text{Max} (M R_{\text{rr}})/L_{\text{dd-vdW}}^3 = |\delta_e|,
\end{eqnarray}
where $\text{Max} (\cdots)$ denotes the maximum matrix element for the dipole-dipole interaction, and $\delta_e$ is the energy defect of the state (97$p_{1/2}$,~99$p_{3/2}$). This gives $L_{\text{dd-vdW}} \approx 9.6\mu$m. In order to implement efficiently the entanglement protocol in the main text, we may set $L > 20\mu$m. In particular, for $L = 26 \mu$m, the matrix $\hat{V}_1$ in Eq.~(\ref{eq101503}) becomes
\begin{eqnarray}
\hat{V}_1^{(97,100)} &=&- \left(\begin{array}{cccc}
 289&0&0&0 \\
 0&	194 & -190 &0  \\
0&-190 & 194 &0  \\
0 &0 &0 &289
 \end{array}
 \right)\text{kHz}.\nonumber
 \end{eqnarray} 
The same analysis for the pair state $n_1=73,n_2=75, l_1=l_2=0$, with relevant energy defects and $R_{\text{rr}}$ listed in Table~\ref{table4}, gives a critical distance $6.1\mu$m. We may set $L=15\mu$m, then the interaction matrix is
\begin{eqnarray}
\hat{V}_1^{(73,75)} &=& \left(\begin{array}{cccc}
 535&0&0&0 \\
 0&	358 & -353 &0  \\
0&-353 & 358 &0  \\
0 &0 &0 &535
 \end{array}
 \right)\text{kHz}.\nonumber
 \end{eqnarray} 

The occurrence of a strong spin-exchange interaction in these cases arises through an interference effect involving a small number of dominant intermediate $p_{1/2}$ and $p_{3/2}$ states. From Table~\ref{table3} and~\ref{table4}, one finds that the strong spin-exchange interaction comes from the dominating scattering channels with energy defects $\delta_2(n_s,n_t)$ and $\delta_3(n_s,n_t)$ in Eq.~(\ref{delta4}). If we only consider the two dominating channels $73p_{1/2}, 74p_{3/2}$ and $73p_{3/2}, 74p_{1/2}$ for two atoms with principal quantum number $73$ and $75$, then $V_+:V_-=1:9$ from Eq.~(\ref{van001}). The channel that contributes next strongly comparing to these two channels is $73p_{3/2}, 74p_{3/2}$, whose energy defect has a sign difference to those of $73p_{1/2}, 74p_{3/2}$ and $73p_{3/2}, 74p_{1/2}$. So, from Eq.~(\ref{van001}), if we consider only this channel with its contribution to the blockade $V_+'$ and $V_-'$, then $V_+':V_-'=-17:-9$, which will reduce $V_+$ furthermore, giving a negligible $V_+$ comparing to $V_-$. For two atoms with principal quantum numbers $97$ and $100$, there are two dominating scattering channels $97p_{1/2}, 99p_{3/2}$ and $97p_{3/2}, 99p_{1/2}$. With only these two channels, we shall have $V_+:V_-=-1:-9$. The channel that contributes next strongly comparing to them is $97p_{1/2}, 99p_{1/2}$. If we consider only this channel, then $V_+'>0=V_-'$, which will reduce $V_+$ furthermore, giving $|V_+/V_-|<1/9$.

\begin{table}
  \begin{tabular}{|c|c||c|c|}
    \hline   atom 1  &  atom 2  & $\delta_e$~(MHz) & $R_{\text{rr}}$~(GHz$\mu \text{m}^3$)\\ \hline
 96$p_{1/2}$&100$p_{1/2}$&-779&     98.3 \\
 96$p_{1/2}$&100$p_{3/2}$ &-685&     96.8 \\
 96$p_{3/2}$ &100$p_{1/2}$&-672&     100.1 \\
 96$p_{3/2}$ &100$p_{3/2}$ &-578&     98.5 \\
 97$p_{1/2}$&99$p_{1/2}$&-62&     98.4 \\
 97$p_{1/2}$&99$p_{3/2}$ &35&     100.2 \\
 97$p_{3/2}$ &99$p_{1/2}$&42&     96.8 \\
 97$p_{3/2}$ &99$p_{3/2}$ &139&     98.6 \\
 98$p_{1/2}$&98$p_{1/2}$&177&     1.7 \\
 98$p_{1/2}$&98$p_{3/2}$ &277&     1.6 \\
 98$p_{3/2}$ &98$p_{1/2}$&277&     1.7 \\
 98$p_{3/2}$ &98$p_{3/2}$ &378&     1.7 \\
 \hline
  \end{tabular}
  \caption{  \label{table3}  Table of relevant scattering channels via dipole-dipole interaction for a pair state with $n_1=97,n_2=100, l_1=l_2=0$.}
\end{table}

\begin{table}
  \begin{tabular}{|c|c||c|c|}
    \hline   atom 1  &  atom 2  & $\delta_e$~(MHz) & $R_{\text{rr}}$~(GHz$\mu \text{m}^3$)\\ \hline
 73$p_{1/2}$&74$p_{3/2}$&-51&30.5\\
 73$p_{3/2}$&74$p_{1/2}$&-41&29.5\\
 73$p_{3/2}$&74$p_{3/2}$&198&30.1\\
 74$p_{1/2}$&73$p_{1/2}$&-291&0.5\\
 74$p_{1/2}$&73$p_{3/2}$&-41&0.5\\
 74$p_{3/2}$&73$p_{1/2}$&-51&0.5\\
 74$p_{3/2}$&73$p_{3/2}$&198&0.5\\
 \hline
  \end{tabular}
  \caption{  \label{table4}  Table of relevant scattering channels via dipole-dipole interaction for a pair state with $n_1=73,n_2=75, l_1=l_2=0$. }
\end{table}

\section{State evolution during Pulse 2}\label{appC}
Pulse 2 consists of a two-photon excitation of atom B with effective Rabi frequency $\Omega^{\text{B} }$. This pulse excites $|\uparrow\rangle$ to $|\Downarrow\rangle$ via the $6P$ state as in Fig.~1(b). By writing the state vector as $\alpha_{\overline{r}\overline{g}} |\Uparrow \uparrow\rangle + \alpha_{r_+} |r_{+}\rangle + \alpha_{r_-} |r_{-}\rangle$, the time evolution is given by
\begin{eqnarray}
 i\frac{d}{dt}\left( \begin{array}{c} \alpha_{\overline{r}\overline{g}} \\ \alpha_{r_+} \\ \alpha_{r_-}\end{array}
\right)  &=& \left( \begin{array}{ccc} 0&\Omega_1  & \Omega_1 \\
\Omega_1 & V_+ &0\\
\Omega_1   &0& V_-
\end{array}
\right)\left( \begin{array}{c} \alpha_{\overline{r}\overline{g}}\\\alpha_{r_+}\\ \alpha_{r_-}\end{array}
\right),
\end{eqnarray}
where $\Omega_1 = \Omega^{\text{B} }/(2\sqrt2)$. In the equation above, we have $|V_-| \gg |V_+|, \Omega_1$. Then at the time
\begin{eqnarray}
&&\tau_2 = \frac{\pi}{\sqrt{V_+^2+4 \Omega_1 ^2-\frac{4 \Omega_1 ^3}{V_-} } } \approx \frac{\pi}{2 \Omega_1 } \left [  1- \frac{ V_+^2}{8 \Omega_1 ^2 }  \right],
\end{eqnarray}
the population on the Rydberg Bell state is maximal
\begin{eqnarray}
|{\alpha}_{r_+}|&=&\frac{2 \Omega_1  }{\sqrt{V_+^2+4 \Omega_1 ^2-\frac{4 \Omega_1 ^3}{V_-}  }}\approx  1- \frac{ V_+^2}{ [ \Omega^{\text{B} }] ^2 }  .
\end{eqnarray}

\section{State evolution during Pulse 3}\label{appD}
\begin{figure}
\includegraphics[width=3.3in]
{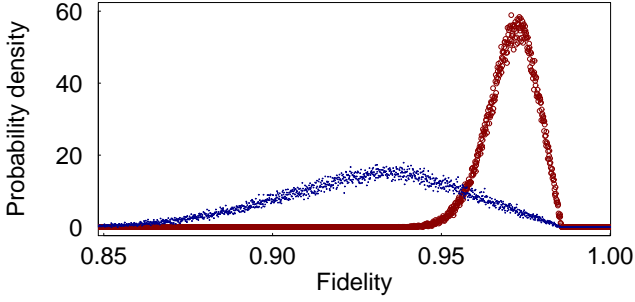}
 \caption{(Color online) Probability distribution of fidelity for the ground Bell state when each of the four two-photon Rabi frequencies in pulse 3 obeys a uniform distribution in $[1-\epsilon,~1+\epsilon]\Omega$, where $\epsilon=0.1~(0.2)$ for the empty circle and dot.  \label{fig004} } 
\end{figure}

As analyzed in the main text, there are 8 states relevant to the dynamics of pulse 3. We write the wavefunction for the system as
 \begin{eqnarray}
|\psi\rangle &=&\alpha_{\underline{g}\overline{g}} |m_F^{(\text{A})}=-1;m_F^{(\text{B})}=2   \rangle\nonumber\\
 &&  + \alpha_{\overline{g}\underline{g}} |m_F^{(\text{A})}=2;m_F^{(\text{B})}=-1   \rangle\nonumber\\
 &&  +\alpha_{\underline{g}\underline{r}} |m_F^{(\text{A})}=-1; R_-^{(\text{B})} \rangle +\alpha_{\overline{g}\overline{r}} |m_F^{(\text{A})}=2; R_+^{(\text{B})} \rangle \nonumber\\
 &&   +\alpha_{\underline{r}\underline{g}} | R_-^{(\text{A})} ;m_F^{(\text{B})}=-1\rangle +\alpha_{\overline{r}\overline{g}} |R_+^{(\text{A})} ; m_F^{(\text{B})}=2 \rangle \nonumber\\
 &&  + \alpha_{\overline{r}\underline{r}} |R_+^{(\text{A})}; R_-^{(\text{B})} \rangle +\alpha_{\underline{r}\overline{r}} |R_-^{(\text{A})}; R_+^{(\text{B})}  \rangle,
\end{eqnarray}
where $|R_\pm^{(\alpha)}\rangle =|m_J^{(\alpha)}m_I^{(\alpha)}=\frac{\pm 1}{2}\frac{1}{2}  \rangle$, then the dynamics is determined by the Schr\"{o}dinger equation,
\begin{widetext}
\begin{eqnarray}
i\frac{d}{dt}\left(\begin{array}{c}
\alpha_{\underline{g}\overline{g}}  \\\alpha_{\overline{g}\underline{g}} \\ \alpha_{\underline{g}\underline{r}} \\ \alpha_{\overline{g}\overline{r}}    \\\alpha_{\underline{r}\underline{g}}  \\
 \alpha_{\overline{r}\overline{g}}   \\ \alpha_{\underline{r}\overline{r}}    \\\alpha_{\overline{r}\underline{r}} \end{array} \right)
&=& \frac{1}{2}\left(\begin{array}{cccccccc}
0&0&   \Omega^{\text{B}} e^{i\phi^{\text{B}}} & 0&0&   \Omega_{\text{A}} e^{i\phi_{\text{A}}} & 0&0\\
0&0&  0&  \Omega_{\text{B}}  e^{i\phi_{\text{B}}}& \Omega^{\text{A}} e^{i\phi^{\text{A}}}&0&   0&0\\
 \Omega^{\text{B}} e^{-i\phi^{\text{B}}} &0&0& 0&0&0&0& \Omega_{\text{A}}  e^{i\phi_{\text{A}}}   \\
0& \Omega_{\text{B}} e^{-i\phi_{\text{B}}}&0& 0&0&0& \Omega^{\text{A}}  e^{i\phi^{\text{A}}}&0   \\
0& \Omega^{\text{A}} e^{-i\phi^{\text{A}}} &0& 0&0&0& \Omega_{\text{B}} e^{i\phi_{\text{B}}}&0   \\
\Omega_{\text{A}} e^{-i\phi_{\text{A}}} &0& 0& 0&0&0&0 & \Omega^{\text{B}}  e^{i\phi^{\text{B}}} \\
0& 0& 0&\Omega^{\text{A}}  e^{-i\phi^{\text{A}}} & \Omega_{\text{B}}  e^{-i\phi_{\text{B}}} &0 & 2V_{\text{s}}  &2V_{\text{c}} \\
0& 0& \Omega_{\text{A}}  e^{-i\phi_{\text{A}}}  &0&0& \Omega^{\text{B}}  e^{-i\phi^{\text{B}}}  &2V_{\text{c}} &2 V_{\text{s}}
 \end{array} \right)
\left(\begin{array}{c}
\alpha_{\underline{g}\overline{g}}  \\\alpha_{\overline{g}\underline{g}} \\ \alpha_{\underline{g}\underline{r}} \\ \alpha_{\overline{g}\overline{r}}    \\\alpha_{\underline{r}\underline{g}}  \\
 \alpha_{\overline{r}\overline{g}}   \\ \alpha_{\underline{r}\overline{r}}    \\\alpha_{\overline{r}\underline{r}} \end{array} \right),\nonumber\\ \label{Hpulse3}
\end{eqnarray}
\end{widetext}
where $V_{\text{s}} =C_6 /L^6$ and $V_{\text{c}} =C_6^{ex}/L^6$.
The Hamiltonian $\hat{H}_{\text{Pulse3}}$ gives the following coupling
\begin{eqnarray}
\langle \Uparrow  \mathcal{B}|\hat{H}_{\text{Pulse3}} | \downarrow \mathcal{B}\rangle&=& \Omega_{\text{A} } e^{-i\phi_{\text{A} }},\nonumber\\
\langle \Downarrow  \mathcal{B}|\hat{H}_{\text{Pulse3}} | \uparrow \mathcal{B}\rangle&=& \Omega^{\text{A} } e^{-i\phi^{\text{A} }},\nonumber\\
\langle  \mathcal{A}\Uparrow |\hat{H}_{\text{Pulse3}} | \mathcal{A}\downarrow \rangle&=&  \Omega_{\text{B} } e^{-i\phi_{\text{B} }},\nonumber\\
\langle  \mathcal{A}\Downarrow |\hat{H}_{\text{Pulse3}} | \mathcal{A}\uparrow \rangle&=&\Omega^{\text{B} } e^{-i\phi^{\text{B} }}, \label{eq071}
\end{eqnarray}
where the phase factors have been defined previously and $\mathcal{A}$ and $\mathcal{B}$ represent an arbitrary state of atom $A$ and $B$. With all the optical excitation channels active, the Rydberg state can be mapped down to the intermediate states $|$$\Uparrow \uparrow\rangle, |$$\Downarrow\downarrow\rangle$, $|$$\uparrow\Uparrow \rangle$, and $|$$\downarrow\Downarrow\rangle$, where for convenience we introduce new basis states
\begin{eqnarray}
|\mathscr{U}_\pm\rangle &\equiv &\left[ e^{i\phi^{\text{B} }  } | \Uparrow \uparrow\rangle \pm e^{i\phi^{\text{A} }} |\uparrow\Uparrow  \rangle \right]/\sqrt2,\nonumber\\
|\mathscr{D}_\pm\rangle &\equiv &\left[ e^{i\phi_{\text{B} }} | \Downarrow\downarrow \rangle\pm  e^{i\phi_{\text{A} }} |\downarrow \Downarrow\rangle\right]/\sqrt2.
\end{eqnarray}
Then it is easily seen that only states $|\mathscr{U}_+\rangle$ and $|\mathscr{D}_+\rangle$ are coupled with entangled Rydberg state $|r_{+}\rangle$, while the other two states $|\mathscr{U}_-\rangle$ and $|\mathscr{D}_-\rangle$ are decoupled from $|r_{+}\rangle$ when the Rabi frequencies satisfy $\Omega\equiv \Omega_{\text{A} }= \Omega^{\text{A} } = \Omega_{\text{B} } =\Omega^{\text{B} }$. The states $|\mathscr{U}_+\rangle$ and $|\mathscr{D}_+\rangle$ are further coupled to the ground states $|$$\downarrow\uparrow\rangle$ and $|$$\uparrow\downarrow\rangle$. If we define the basis states
\begin{eqnarray}
|\Phi_\pm\rangle &\equiv &\left[ e^{i(\phi_{\text{A} } +  \phi^{\text{B} } ) } | \downarrow \uparrow\rangle \pm e^{i(\phi^{\text{A} } +  \phi_{\text{B} } ) } |\uparrow\downarrow  \rangle \right]/\sqrt2,
\end{eqnarray}
it is easy to show that only the state $|\Phi_{+}\rangle $ is coupled with $|\mathscr{U}_+\rangle$ and $|\mathscr{D}_+\rangle$. Hence only four states $|r_{+}\rangle,~|\mathscr{U}_+\rangle,~|\mathscr{D}_+\rangle $ and $|\Phi_{+}\rangle $ interact during pulse 3. When atom $A$ is excited by four sets of light fields, each in the diamond-shaped configuration, we can have $\phi_{\text{A} } = \phi^{\text{A} }$; similarly for atom $B$. In this case, $|$$g_\pm\rangle = |\Phi_\pm\rangle$, up to overall phase, and we create a ground Bell state $|$$g_+\rangle$. We henceforth assume that $\phi_{\text{A} } = \phi^{\text{A} }$ and $\phi_{\text{B} } = \phi^{\text{B} }$ are satisfied, which justifies neglecting these phases in the main text for simplicity.
The main practical difficulty is to set all four Rabi frequencies in Eq.~(\ref{Hpulse3}) or~(\ref{eq071}) equal to each other. In order to show that the method proposed above is robust against dispersion in the Rabi frequencies, we numerically simulate the system evolution by independently varying $\Omega_{\text{A} }, \Omega^{\text{A} },\Omega_{\text{B} }$ and $\Omega^{\text{B} }$ among the interval $[1-\epsilon,~1+\epsilon]\Omega$~(i.e., each Rabi frequency obeys a uniform distribution) for pulse 3, with $\Omega=\sqrt{V_+V_-}$, $\tau_2=(\sqrt2\pi/\Omega) \left [ 1 -  (V_+/ \Omega) ^2  \right]$, and $\tau_3=\pi/\Omega$. By performing $10^5$ such simulations, we find that almost all fidelities are larger than $95\%$~($85\%$) for $\epsilon=0.1~(0.2)$, as shown in Fig.~\ref{fig004}. We note that an almost identical robustness holds for the fidelity of preparation of $|\Phi_+\rangle,$ as opposed to $|g_+\rangle$, in the case where the phases $\phi_{\text{A} }   ,\phi^{\text{A} },\phi_{\text{B} }$ and $\phi^{\text{B} }$ in Eq.~(\ref{eq071}) are chosen randomly.

\section{Entanglement of atomic chains}\label{appE}
In this section we describe in detail the entanglement of a one-dimensional chain of $4$ and $6$ atoms separated by a distance $L = 15 \mu$m. We consider the following Rydberg target states: $n_1=73, l_1=0$ and $n_2=75, l_2=0$.  The relevant Rydberg interaction between two nearest-neighboring atoms is characterized by the spin-exchange process:
\begin{eqnarray}
V_+/(2\pi) \approx 5 \text{kHz},~~
V_-/(2\pi) \approx 700\text{kHz}.
\end{eqnarray}
The strongest blockade interaction for non-neighboring atoms is 
\begin{eqnarray}
V_1^{(73,75)} (1,1)/(2\pi)=V_1^{(73,75)} (4,4)/(2\pi) \approx 0.7 \text{kHz}.
\end{eqnarray}
This means we can safely ignore the blockade effect due to multiple Rydberg excitations in the chain. Since the first and second steps in the entangling protocol shown in the main text is simply the entanglement between two nearest neighbors, here we focus on the third and fourth steps.

\subsection{Entanglement among 4 atoms}
Here we consider a system consisting of only four atoms
\begin{eqnarray}
A_1,~~B_1,~~C_1,~~D_1,\nonumber
\end{eqnarray}
as an example. The pumping for step 1 and 2 follow as
\begin{eqnarray}A_1,~~B_1&& \xrightarrow{\text{Step 1: pairwise entanglement} } | g_{+} \rangle ,\nonumber\\
C_1,~~D_1  &&  \xrightarrow{\text{Step 2: pairwise entanglement }} | g_{+} \rangle ,\nonumber
\end{eqnarray}
while step 3 is
\begin{eqnarray}
B_1,~~C_1  &&  \xrightarrow{\text{Step 3: SWAP-gate} }  B_1,~~C_1 \text{ become entangled}.\nonumber
\end{eqnarray}
Here pulse $1,2,3$ and $4,5,6$ are exactly the same as the three pulses in the main text. To illustrate the physics in step 3, we first write out the state after step 2~(here we assume perfect creation of ground Bell state for step 1 and 2),
\begin{eqnarray}
\frac{1}{2}\left( |\uparrow \downarrow\rangle+|\downarrow\uparrow \rangle  \right)_{ A_1,B_1 }\otimes\left( |\uparrow \downarrow\rangle+|\downarrow\uparrow \rangle  \right)_{ C_1,D_1 },\nonumber
\end{eqnarray}
then the two $\pi$ pulses and $2\pi$ pulse of step 3 have the effect
\begin{widetext}
\begin{eqnarray}
&&\text{Two two-photon Rabi processes via 5P, 6P states of atom }C_1,~~T_1 = \frac{\pi}{\Omega},\text{reaching n = 73 Rydberg state} \nonumber\\
\xrightarrow{\text{pulse 7}}&&\frac{1}{2}\left( |\uparrow \downarrow\rangle+|\downarrow\uparrow \rangle  \right)_{ A_1,B_1 }\otimes\left( |\Downarrow \downarrow\rangle+|\Uparrow\uparrow \rangle  \right)_{ C_1,D_1 }=\frac{1}{2}\left(  |\uparrow \downarrow\Uparrow\uparrow\rangle+|\downarrow\uparrow\Downarrow \downarrow \rangle   +   |\uparrow \downarrow\Downarrow\downarrow\rangle+|\downarrow\uparrow\Uparrow \uparrow \rangle    \right)_{ A_1,B_1 , C_1,D_1 },\nonumber\\
&&\text{Two two-photon Rabi processes via 5P, 6P states of atom }B_1,~~T_2 \approx \frac{2\pi}{\Omega},\text{using n = 75 Rydberg state} \nonumber\\
\xrightarrow{\text{pulse 8}}&&\frac{1}{2}\left( |\uparrow \downarrow\Uparrow\uparrow\rangle+|\downarrow\uparrow\Downarrow \downarrow \rangle -  |\uparrow   \uparrow\Uparrow  \downarrow\rangle-  |\downarrow \downarrow\Downarrow \uparrow\rangle\right)_{ A_1,B_1 , C_1,D_1 } ,\nonumber\\
&&\text{Two two-photon Rabi processes via 5P, 6P states of atom }C_1,~~T_3 = \frac{\pi}{\Omega},\text{depopulating n = 73 Rydberg state} \nonumber\\
\xrightarrow{\text{pulse 9}}&&\frac{1}{2}\left( |\uparrow \downarrow\downarrow\uparrow\rangle+|\downarrow\uparrow\uparrow \downarrow \rangle  - |\uparrow   \uparrow\downarrow  \downarrow\rangle-  |\downarrow \downarrow \uparrow\uparrow\rangle\right)_{ A_1,B_1 , C_1,D_1 }. \label{4atomentanglement}
\end{eqnarray}
\end{widetext}
where we assume that the state $|\downarrow\Uparrow\rangle_{B_1,C_1}$ can not be excited to state $|\Uparrow\Uparrow\rangle_{B_1,C_1}$~(similar for the other state $|\uparrow\Downarrow\rangle_{B_1,C_1}$) during pulse 8, simply because~(the value is for two nearest-neighboring atoms)
\begin{eqnarray}
V_1^{(73,75)} (1,1)/(2\pi)=V_1^{(73,75)} (4,4)/(2\pi) \approx 0.5 \text{MHz}, \nonumber
\end{eqnarray}
 is much larger than any other rates in the model. Here we have labeled the two Rabi frequencies $\Omega_{\text{A} } ,\Omega^{\text{A} } $ during pulse 7 and 9, and $\Omega_{\text{B} } ,\Omega^{\text{B} } $ during pulse 8 as $\Omega$. Pulse 7 and 9 are simply $\pi$ pulses. To understand the $2\pi$ pulse~(i.e., pulse 8) above, let us write the Hamiltonian
\begin{eqnarray}
 \hat{H}_{\text{Pulse 8}}^{(1)}\!  &\!=\!& \!  \frac{1 }{2\sqrt2}\left( \begin{array}{cccc} 0& 0&\Omega e^{i\phi }& -\Omega e^{i\phi }  \\ 0& 0&\Omega e^{i\phi }& \Omega e^{i\phi }  \\
\Omega  e^{-i\phi } &\Omega  e^{-i\phi }  &2\sqrt2 V_+ &0\\
-\Omega  e^{-i\phi }  &\Omega  e^{-i\phi }  &0 &2\sqrt2 V_-
\end{array}
\right),\nonumber\\
&&\label{pulse8H}
\end{eqnarray}
where the basis vectors are $|$$\uparrow\Uparrow \rangle_{ B_1,C_1 }$, $|$$\downarrow \Downarrow \rangle_{ B_1,C_1 }$, $|r_{+}\rangle_{ B_1,C_1 }$ and $|r_{-}\rangle_{ B_1,C_1 }$, and $\phi $ is similar to $\phi_{\text{B} }=  \phi^{\text{B} }$ in the main text. To have an analytical result, we neglect the population on $|r_{-}\rangle_{ B_1,C_1 }$, and also approximate $V_+\approx0$, then we have
\begin{eqnarray}
|\downarrow\Downarrow\rangle \xrightarrow{2\pi ~\text{pulse with Rabi frequency}~\Omega }- |\uparrow\Uparrow\rangle ,
\nonumber \\
|\uparrow\Uparrow\rangle\xrightarrow{2\pi ~\text{pulse with Rabi frequency}~\Omega }- |\downarrow\Downarrow\rangle.
\end{eqnarray}

From Eq.~(\ref{4atomentanglement}), it is clear that in the computation basis $|\uparrow \uparrow\rangle,~~  |\uparrow \downarrow\rangle,~~|\downarrow\uparrow \rangle,~~  |\downarrow\downarrow\rangle $, we have simply realized a SWAP-like gate that can be represented by the matrix~(taking into account the $\pi$-phase on each basis state due to optical excitation of pulse 7 and 9 on atom $C_1$)
\begin{eqnarray}
\overline{\text{SWAP}}& =& \left( \begin{array} {cccc}-1 & 0& 0 &0 \\
0 & 0 &1 &0 \\
0 & 1 &0 &0 \\
0 & 0 &0 &-1 \end{array}
\right),
\end{eqnarray}
which is very similar to a SWAP gate, up to a $\pi$ phase change on the basis states $|\uparrow \downarrow\rangle,~~|\downarrow\uparrow \rangle $ relative to the basis states $|\uparrow \uparrow\rangle,~~  |\downarrow\downarrow\rangle$. 

\begin{figure}
\includegraphics[width=3in]
{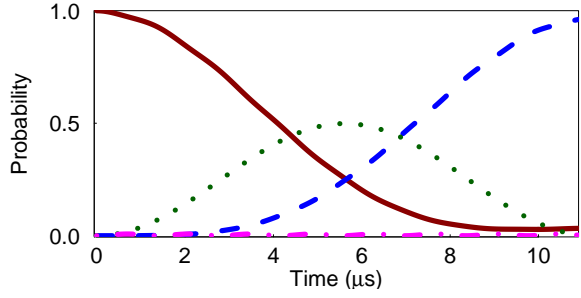}
 \caption{(Color online) State evolution starting from $|\uparrow\Uparrow\rangle$. The (red) solid, (blue) dashed, (green) dotted, and (magenta) dash-dotted curves represent population on state $|$$\uparrow\Uparrow \rangle_{ B_1,C_1 }$, $|$$\downarrow \Downarrow \rangle_{ B_1,C_1 }$, $|r_{+}\rangle_{ B_1,C_1 }$ and $|r_{-}\rangle_{ B_1,C_1 }$, respectively. Here $\Omega=89$kHz. At the end of pulse duration $t_{pulse8}=11.12\mu s$~(this duration is $0.12\mu s$ shorter than $2\pi/\Omega$), $|\langle \Psi|\downarrow \Downarrow \rangle_{ B_1,C_1 }|^2=96.649\%$. \label{figS4}}
\end{figure}

Here, the Rabi frequencies in pulse 7 and pulse 9 are only limited by the available laser, since there is one~(approximately one) Rydberg excitation during pulse 7~(9). However, we shall use typical Rabi frequency $\sqrt{V_+V_-}$ during pulse 8. By choosing $\Omega=89$kHz and pulse duration $t_{pulse8}=11.12\mu s$, we numerically solve the state evolution by Eq.~(\ref{pulse8H}) and find that pulse 8 can transform from state $|$$ \uparrow\Uparrow\rangle$ to state $|$$\downarrow \Downarrow \rangle_{ B_1,C_1 }$ with a fidelity $\mathcal{F}_{\downarrow \uparrow}=96.649\%$~(the remaining population is mainly in $|\downarrow\Downarrow\rangle$), as shown in Fig.~\ref{figS4}. Following pulse 8, pulse 9 shall be a $\pi$ pulse transferring $|$$\downarrow \Downarrow \rangle_{ B_1,C_1 }$ back to $|\uparrow \downarrow\rangle$, although pulse 9 will also transfer $\sim\sqrt{0.033}|\downarrow\Downarrow\rangle$ back to and up from $|\downarrow\uparrow\rangle$. Neglecting population on $|$$\downarrow \Downarrow \rangle_{ B_1,C_1 }$, then pulses 7,8 and 9 have the following property,
\begin{eqnarray}
&&|\downarrow \uparrow \rangle\xrightarrow[\text{atom} C_1]{\pi}  -i |\downarrow\Downarrow\rangle \xrightarrow[\text{atom} B_1]{2\pi}i |\uparrow\Uparrow\rangle \xrightarrow[\text{atom} C_1]{\pi}|\uparrow \downarrow\rangle,
\nonumber \\
&&|\uparrow \downarrow\rangle\xrightarrow[\text{atom} C_1]{\pi}-i|\uparrow\Uparrow\rangle\xrightarrow[\text{atom} B_1]{2\pi} i |\downarrow\Downarrow\rangle \xrightarrow[\text{atom} C_1]{\pi}|\downarrow\uparrow\rangle ,\nonumber
\end{eqnarray}
where pulse 7 and 9 can be ideal $\pi$ pulses. To find the fidelity of the gate through pulse 7 to pulse 9, we shall also study state evolution of $|\uparrow \uparrow\rangle,~~  |\downarrow\downarrow\rangle $. Choosing initial state $|\uparrow \uparrow\rangle$, then pulse 8 have the Hamiltonian beginning with state $-i|\uparrow \Downarrow\rangle$
\begin{eqnarray}
 \hat{H}_{\text{Pulse 8}}^{(2)}\!  &\!=\!& \!  \left( \begin{array}{cc} 0&\frac{1 }{2}\Omega e^{i\phi(B_1)}  \\\frac{1 }{2} \Omega e^{i\phi(B_1)} & V_{73,75} (4,4)
\end{array}
\right),\label{pulse8H02}
\end{eqnarray} 
which gives us a population of $\mathcal{F}_{\uparrow\uparrow}=99.976\%$ on state $|\uparrow \Downarrow\rangle$ at the end of pulse 8 with $\Omega=89$kHz. Note that pulse 9 can ideally transfer this state back to $|\uparrow \uparrow\rangle$ in MHz rate. Then the fidelity of the SWAP-like gate is
\begin{eqnarray}
\mathcal{F}_{\mathrm{SWAP}}\equiv \frac{1}{4} (\mathcal{F}_{\uparrow\downarrow } +\mathcal{F}_{\downarrow \uparrow} +\mathcal{F}_{\uparrow\uparrow} +\mathcal{F}_{\downarrow \downarrow } )=98.31\%.
\end{eqnarray}

\subsection{Entanglement among 6 atoms}
A minimal model to illustrate global entanglement of $4N$ atoms can be shown by a system consisting of only six atoms. In order to characterize the entanglement, we consider a system consisting of the following six atoms
\begin{eqnarray}
A_1,~~B_1,~~C_1,~~D_1,~~A_2,~~B_2,
\end{eqnarray}
as an example. The pumping for step 1 and 2 follows as
\begin{eqnarray}A_1,~B_1;~A_2,~B_2 && \xrightarrow{\text{Step 1: pairwise entanglement} } | g_{+} \rangle\otimes| g_{+} \rangle ,\nonumber\\
C_1,~D_1  &&  \xrightarrow{\text{Step 2: pairwise entanglement} } | g_{+} \rangle ,\nonumber
\end{eqnarray}
while step 3 and step 4 are
\begin{eqnarray}
B_1,~~C_1  &&  \xrightarrow{\text{Step 3: SWAP-gate } }  B_1,~~C_1 \text{ become entangled},\nonumber\\
D_1,~~A_2  &&  \xrightarrow{\text{Step 4: SWAP-gate  } } D_1,~~A_2  \text{ become entangled}.\nonumber
\end{eqnarray}
After step 3, we have a state in Eq.~(\ref{4atomentanglement}) for $A_1,~~B_1,~~C_1,~~D_1$. To illustrate the physics in step 4, we first write out the state after step 3~(here we assume perfect creation of ground Bell state for step 1 and 2, and entanglement in step 3),
\begin{eqnarray}
&&\frac{1}{2\sqrt2}  \left( |\uparrow \downarrow\downarrow\uparrow\rangle+|\downarrow\uparrow\uparrow \downarrow \rangle  - |\uparrow   \uparrow\downarrow  \downarrow\rangle-  |\downarrow \downarrow \uparrow\uparrow\rangle\right)_{ A_1,B_1 , C_1,D_1 } 
\nonumber\\ &&~~\otimes \left( |\uparrow \downarrow\rangle+|\downarrow\uparrow \rangle  \right)_{ A_2,B_2 }. \nonumber
\end{eqnarray}
The pulse in this final step is exactly the same as in Eq.~(\ref{4atomentanglement}).
\begin{widetext}
\begin{eqnarray}
&&\text{Two two-photon Rabi processes via 5P, 6P states of atom }A_2,~~T_1 = \frac{\pi}{\Omega},\text{reaching n = 73 Rydberg state} \nonumber\\
\xrightarrow{\text{pulse 10}}&&  \frac{1}{2\sqrt2}  \left[ \left( |\uparrow \downarrow\downarrow\rangle-  |\downarrow \downarrow \uparrow\rangle\right)| \uparrow \rangle+ \left( |\downarrow\uparrow\uparrow \rangle  - |\uparrow   \uparrow\downarrow \rangle\right)|  \downarrow\rangle   \right]_{ A_1,B_1 , C_1,D_1 } \otimes\left( |\Downarrow \downarrow\rangle+|\Uparrow\uparrow \rangle  \right)_{ A_2,B_2 } \nonumber\\
&& =  \frac{1}{2\sqrt2}  \left[ \left( |\uparrow \downarrow\downarrow\rangle-  |\downarrow \downarrow \uparrow\rangle\right)| \uparrow\Downarrow \downarrow \rangle+ \left( |\downarrow\uparrow\uparrow \rangle  - |\uparrow   \uparrow\downarrow \rangle\right)|  \downarrow\Downarrow \downarrow\rangle + \left( |\uparrow \downarrow\downarrow\rangle-  |\downarrow \downarrow \uparrow\rangle\right)| \uparrow \Uparrow\uparrow\rangle+ \left( |\downarrow\uparrow\uparrow \rangle  - |\uparrow   \uparrow\downarrow \rangle\right)|  \downarrow\Uparrow\uparrow\rangle    \right]  ,\nonumber\\
&&\text{Two two-photon Rabi processes via 5P, 6P states of atom }D_1,~~T_2 \approx \frac{2\pi}{\Omega},\text{using n = 75 Rydberg state} \nonumber\\
\xrightarrow{\text{pulse 11}}&&  \frac{1}{2\sqrt2}  \left[ \left( |\uparrow \downarrow\downarrow\rangle-  |\downarrow \downarrow \uparrow\rangle\right)| \uparrow\Downarrow \downarrow \rangle
- \left( |\downarrow\uparrow\uparrow \rangle  - |\uparrow   \uparrow\downarrow \rangle\right)|  \uparrow \Uparrow \downarrow\rangle
- \left( |\uparrow \downarrow\downarrow\rangle-  |\downarrow \downarrow \uparrow\rangle\right)|\downarrow\Downarrow \uparrow\rangle
+ \left( |\downarrow\uparrow\uparrow \rangle  - |\uparrow   \uparrow\downarrow \rangle\right)|  \downarrow\Uparrow\uparrow\rangle    \right]    ,\nonumber\\
&&\text{Two two-photon Rabi processes via 5P, 6P states of atom }A_2,~~T_3 = \frac{\pi}{\Omega},\text{depopulating n = 73 Rydberg state} \nonumber\\
\xrightarrow{\text{pulse 12}}&&  \frac{1}{2\sqrt2}  \left[ \left( |\uparrow \downarrow\downarrow\rangle-  |\downarrow \downarrow \uparrow\rangle\right)| \uparrow\uparrow \downarrow \rangle
- \left( |\downarrow\uparrow\uparrow \rangle  - |\uparrow   \uparrow\downarrow \rangle\right)|  \uparrow \downarrow \downarrow\rangle
- \left( |\uparrow \downarrow\downarrow\rangle-  |\downarrow \downarrow \uparrow\rangle\right)|\downarrow\uparrow \uparrow\rangle
+ \left( |\downarrow\uparrow\uparrow \rangle  - |\uparrow   \uparrow\downarrow \rangle\right)|  \downarrow\downarrow\uparrow\rangle    \right] \nonumber\\
&&=  \frac{1}{2\sqrt2}  \left[ \left( |\uparrow \downarrow\downarrow\rangle-  |\downarrow \downarrow \uparrow\rangle\right)_{ A_1,B_1 , C_1}\left( | \uparrow\uparrow \downarrow \rangle-|\downarrow\uparrow \uparrow\rangle\right)_{ D_1,A_2,B_2 }
+ \left( |\downarrow\uparrow\uparrow \rangle  - |\uparrow   \uparrow\downarrow \rangle\right)_{ A_1,B_1 , C_1} \left( |  \downarrow\downarrow\uparrow\rangle -  |  \uparrow \downarrow \downarrow\rangle \right)_{ D_1,A_2,B_2 } \right] . \nonumber
\end{eqnarray}
\end{widetext}
\bibliography{R_entanglement}

\end{document}